\documentclass[aps,prd,showpacs,notitlepage,nofootinbib,preprintnumbers,amsmath,amssymb]{revtex4-1}

\usepackage{graphics,graphicx}% Include figure files
\usepackage{dcolumn}% Align table columns on decimal point
\usepackage{bm}% bold math
\usepackage{epsfig}
\usepackage[usenames]{color}
\usepackage{hyperref} 
\usepackage{mathbbol}
\usepackage{epstopdf}
\usepackage{simplewick}
\usepackage[utf8]{inputenc} 
\usepackage{slashed}
\usepackage{stackrel}

\def\eq#1{{Eq.~(\ref{#1})}}
\def\fig#1{{Fig.~\ref{#1}}}
\newcommand{\ben}{\begin{eqnarray*}}
\newcommand{\een}{\end{eqnarray*}}
\newcommand{\un}[1]{\underline{#1}}

\newcommand{\tr}{\mbox{tr}}
\newcommand{\thalf}{\tfrac{1}{2}}
\newcommand{\llangle}{\Big\langle \!\! \Big\langle}
\newcommand{\rrangle}{\Big\rangle \!\! \Big\rangle}

\newcommand{\as}{\alpha_s}

\newcommand{\dhd}{{\textstyle d}
\lower.03ex\hbox{\kern-0.38em$^{\scriptstyle-}$}\kern-0.05em{}}
\newcommand{\dbar}{{\textstyle \delta}
\lower.03ex\hbox{\kern-0.38em$^{\scriptstyle-}$}\kern-0.05em{}}
\newcommand{\half}{{1\over 2}}

\newcommand{\bra}[1]{\left\langle #1 \right|}
\newcommand{\ket}[1]{\left| #1 \right\rangle}

\newcommand{\ul}[1]{\underline{#1}}

\newcommand{\Tr}{\mathrm{Tr}}

\newcommand{\cc}{\mbox{c.c.}}

\newcommand{\oone}{
\begin{picture}(10,8)
\put(5,5){\circle{8}}
\put(2.9,2.5){{\scriptsize 1}}
\end{picture}
}
\newcommand{\otwo}{
\begin{picture}(10,8)
\put(5,5){\circle{8}}
\put(2.9,2.5){{\scriptsize 2}}
\end{picture}
}
\newcommand{\othree}{
\begin{picture}(10,8)
\put(5,5){\circle{8}}
\put(2.9,2.5){{\scriptsize 3}}
\end{picture}
}

\begin{document}

\title{Small-$x$ Helicity Evolution: an Operator Treatment} 

\author{Yuri V. Kovchegov} 
         \email[Email: ]{kovchegov.1@osu.edu}
         \affiliation{Department of Physics, The Ohio State
           University, Columbus, OH 43210, USA}

\author{Matthew D. Sievert}
  \email[Email: ]{sievertmd@lanl.gov}
	\affiliation{Theoretical Division, Los Alamos National Laboratory, Los Alamos, NM 87545, USA}

\begin{abstract}
We rederive the small-$x$ evolution equations governing quark helicity distribution in a proton using solely an operator-based approach. In our previous works on the subject, the evolution equations were derived using a mix of diagrammatic and operator-based methods. In this work, we re-derive the double-logarithmic small-$x$ evolution equations for quark helicity in terms of the ``polarized Wilson lines", the operators consisting of light-cone Wilson lines with one or two non-eikonal local operator insertions which bring in helicity dependence. For the first time we give explicit and complete expressions for the quark and gluon polarized Wilson line operators, including insertions of both the gluon and quark sub-eikonal operators. We show that the double-logarithmic small-$x$ evolution of the ``polarized dipole amplitude" operators, made out of regular light-cone Wilson lines along with the polarized ones constructed here, reproduces the equations derived in our earlier works. The method we present here can be used as a template for determining the small-$x$ asymptotics of any transverse momentum-dependent (TMD) quark (or gluon) parton distribution functions (PDFs), and is not limited to helicity. 
\end{abstract}


\maketitle

\tableofcontents

%||||||||||||||||||||||||||||||||||||||||||||||||||||||||||||||||||||||||||||||||||||||||||||||||||||||||||||||||||||||||
%||||||||||||||||||||||||||||||||||||||||||||||||||||||||||||||||||||||||||||||||||||||||||||||||||||||||||||||||||||||||
%||||||||||||||||||||||||||||||||||||||||||||||||||||||||||||||||||||||||||||||||||||||||||||||||||||||||||||||||||||||||

%%%%%%%%%%%%%%%%%%%%%%%%%%%%%%%%%%%%%%%%%%%%%%%%%%%%%%%%%%%%%%%%%

\section{Introduction}

Understanding the small-$x$ asymptotics of the quark and gluon helicity distributions is very important for the efforts to resolve the proton spin puzzle: the current measured amounts of the proton's spin carried by its quarks and gluons comes up short of $1/2$, the spin of the proton \cite{Accardi:2012qut,Aschenauer:2013woa,Aschenauer:2015eha,Aschenauer:2016our}. On the theoretical side, the helicity sum
rules~\cite{Jaffe:1989jz,Ji:1996ek,Ji:2012sj} require the proton spin carried by the quarks and gluons, along with the orbital angular momentum (OAM) of the quarks and gluons, to add up to $1/2$ (see \cite{Leader:2013jra} for a review). Therefore, the missing spin could be found either in the less well-known gluon helicity PDF, in the quark and gluon OAM or in the small Bjorken $x$ region, whose contribution to the proton polarization has not been explored. Indeed, experimental measurements of the double-longitudinal spin asymmetry $A_{LL}$, which is used to extract the quark and gluon helicity PDFs, are always limited to $x \ge x_{min}$ with $x_{min}$ the smallest value of the Bjorken variable $x$ which a given experiment allows to probe. This way, any given high-energy experiment can never measure the quark and gluon polarizations down to $x=0$: theoretical input appears to be needed to better constrain the amount of quark and gluon spin at small $x$, which, in turn, would help us get a better handle on the proton spin puzzle. 

In recent years, evolution equations describing the quark and gluon helicity distributions at small Bjorken $x$ have been derived in \cite{Kovchegov:2015pbl,Kovchegov:2016weo,Kovchegov:2016zex,Kovchegov:2017jxc,Kovchegov:2017lsr} (see also \cite{Bartels:1995iu,Bartels:1996wc} for earlier calculations based on a different method). These evolution equations were solved in the large-$N_c$ limit (with $N_c$ the number of quark colors), leading to the following $x$-dependence for the quark and gluon helicity PDFs \cite{Kovchegov:2016weo,Kovchegov:2017jxc,Kovchegov:2017lsr} in that limit and at perturbatively small values of the strong coupling constant $\as$ (such that the 't Hooft coupling is small, $\as \, N_c \ll 1$):
\begin{align}\label{qG_helicity}
 \Delta q (x, Q^2) \sim \left( \frac{1}{x} \right)^{\frac{4}{\sqrt{3}} \, \sqrt{\frac{\as
      \, N_c}{2 \pi}}}, \ \ \ \ \ \Delta G (x, Q^2) \sim \left( \frac{1}{x} \right)^{\frac{13}{4 \sqrt{3}} \,
\sqrt{\frac{\as \, N_c}{2 \pi}}}.
\end{align}
The resummation parameter in the equations derived and studied in \cite{Kovchegov:2015pbl,Kovchegov:2016weo,Kovchegov:2016zex,Kovchegov:2017jxc,Kovchegov:2017lsr} was $\as \, \ln^2 (1/x)$. We will refer to the resummation of this parameter as the Double Logarithmic Approximation (DLA). This parameter, originally introduced by Kirschner and Lipatov \cite{Kirschner:1983di}, arises in certain types of small-$x$ evolution describing e.g. polarization or baryon number transfer from larger to smaller $x$ \cite{Kirschner:1983di, Kirschner:1985cb,Kirschner:1994vc,Kirschner:1994rq,Griffiths:1999dj,Itakura:2003jp,Bartels:2003dj}. This parameter does not exist in the more familiar Balitsky--Fadin--Kuraev--Lipatov (BFKL) \cite{Kuraev:1977fs,Balitsky:1978ic} small-$x$ evolution for the unpolarized gluon distribution, which at the leading order resums powers of  $\as \, \ln (1/x)$.

The helicity evolution equations of \cite{Kovchegov:2015pbl,Kovchegov:2016weo,Kovchegov:2016zex,Kovchegov:2017jxc,Kovchegov:2017lsr} were written in the $s$-channel evolution formalism previously used to derive the unpolarized Balitsky--Kovchegov (BK)
\cite{Balitsky:1995ub,Balitsky:1998ya,Kovchegov:1999yj,Kovchegov:1999ua}
and Jalilian-Marian--Iancu--McLerran--Weigert--Leonidov--Kovner
(JIMWLK)
\cite{Jalilian-Marian:1997dw,Jalilian-Marian:1997gr,Iancu:2001ad,Iancu:2000hn}
evolution equations. The helicity evolution was written in terms of the so-called quark or gluon ``polarized Wilson lines", which were defined originally in \cite{Kovchegov:2015pbl} as part of the scattering amplitude of a longitudinally polarized quark or gluon (projectile) on a longitudinally polarized proton (target) which is proportional to the product of the projectile and target polarizations. The operator describing the quark helicity was shown to be related to the ``polarized dipole amplitude": the polarization-dependent part of the scattering amplitude for a color-singlet quark--antiquark pair. The polarized dipole amplitude was shown to be a correlation function of a trace of polarized and regular light-cone Wilson lines. Similar to the case of the unpolarized Balitsky hierarchy \cite{Balitsky:1995ub,Balitsky:1998ya}, the helicity evolution equations do not close in general. Closed equations were obtained in the large-$N_c$ and the large-$N_c \& N_f$ limits \cite{Kovchegov:2015pbl} (with $N_f$ the number of quark flavors). The large-$N_c$ equations for quark helicity were solved in \cite{Kovchegov:2016weo,Kovchegov:2017jxc} ultimately leading to the $\Delta q$ small-$x$ asymptotics shown in \eq{qG_helicity}. Gluon helicity distribution was studied in \cite{Kovchegov:2017lsr}: new relevant operators had to be defined (see also \cite{Hatta:2016aoc}), their evolution equations were constructed and solved in the large-$N_c$ limit, leading to the small-$x$ asymptotics for $\Delta G$ also shown in \eq{qG_helicity}.

However, an explicit form of the polarized Wilson line operators was not derived in \cite{Kovchegov:2015pbl,Kovchegov:2016weo,Kovchegov:2016zex,Kovchegov:2017jxc}. In \cite{Kovchegov:2017lsr}, the first expression for the polarized {\sl quark} Wilson line was written down. It is given below in \eq{eq:Wpol_fund2}, and consists of two semi-infinite light-cone Wilson lines, with a sub-eikonal component $F^{12}$ of the gluon field strength tensor $F^{\mu\nu}$ sandwiched between them. (This $F^{12}$ insertion can be interpreted as arising from ${\vec \mu} \cdot {\vec B} = \mu_z \, B_z = - \mu_z \, F^{12}$, where a quark with the chromo-magnetic dipole moment $\vec \mu$ is traveling through the chromo-magnetic background field $\vec B$.\footnote{We thank Raju Venugopalan for pointing out this interpretation.}) We see that helicity dependence enters as a sub-eikonal operator insertion between the eikonal Wilson lines. This structure of sub-eikonal corrections at small and large $x$ was also obtained in \cite{Altinoluk:2014oxa,Jalilian-Marian:2017ttv,Chirilli:2018kkw}. However, the expression \eqref{eq:Wpol_fund2} corresponds only to a (sub-eikonal) gluon exchange with the target shown in the left panel of \fig{vpol}. An important quark exchange contribution, shown in the right panel of \fig{vpol}, was missing, as it was not needed in the large-$N_c$ limit largely utilized in \cite{Kovchegov:2017lsr}. In addition, the polarized {\sl gluon} Wilson line has never been constructed explicitly. 

Our aim here is to rederive the results of \cite{Kovchegov:2015pbl} for the quark helicity while working entirely in the operator language. That is, we want to construct explicit complete expressions for the quark and gluon polarized Wilson line operators. We then want to ``evolve" these operators toward small $x$, obtaining helicity evolution equations. The benefits of such a calculation are twofold: on the one hand, we would be able to cross-check the results of  \cite{Kovchegov:2015pbl,Kovchegov:2016weo,Kovchegov:2016zex,Kovchegov:2017jxc,Kovchegov:2017lsr}. On the other hand, the operator formalism we are going to develop here can be similarly applied to other TMDs, such as transversity or the Sivers function, to study their small-$x$ asymptotics. Knowing the small-$x$ behavior of various TMDs has a number of useful phenomenological and theoretical implications. Our present work opens the possibility to systematically derive the small-$x$ asymptotics for all the TMDs in the framework of perturbative quantum chromodynamics (QCD), complementing the efforts in \cite{Balitsky:2015qba,Balitsky:2016dgz,Metz:2011wb,Dumitru:2014vka,Boer:2016xqr,Marquet:2017xwy}. In \cite{KStransversity} we will apply this formalism to study the small-$x$ asymptotics of the quark transversity TMD.   

The paper is structured as follows. In Sec.~\ref{sec:TMDs} we will start with the operator definition of the quark helicity TMD and evaluate it for small $x$, obtaining the expression \eqref{TMD22} relating it to the polarized quark dipole amplitude. The expression we obtain is identical to the one used in \cite{Kovchegov:2015pbl,Kovchegov:2016weo,Kovchegov:2016zex}: however, in \cite{Kovchegov:2015pbl} it was derived by calculating the cross section for the semi-inclusive deep inelastic scattering (SIDIS). Relating the SIDIS cross section to the quark helicity TMD at the leading order in the coupling we read off the quark helicity TMD in \cite{Kovchegov:2015pbl}. The calculation in Sec.~\ref{sec:TMDs} provides an independent cross-check of this result and shows that the SIDIS definition of the quark helicity TMD used in \cite{Kovchegov:2015pbl} is equivalent to the standard operator definition of the same quantity. 

Explicit operator expressions for the polarized quark and gluon Wilson lines are constructed respectively in Subsections \ref{sec:polWfund} and  \ref{sec:polWadj} of Sec.~\ref{sec:polW}. The results are given by \eq{eq:Wpol_all}  for the quarks and by \eq{M:UpolFull} for the gluons. We proceed by constructing the large-$N_c$ evolution equations for ``polarized dipoles" in Sec.~\ref{sec:largeNc} and the large-$N_c \& N_f$ evolution equations in Sec.~\ref{sec:largeNcNf}. The equations are identical to those derived originally in \cite{Kovchegov:2015pbl}. 

We conclude in Sec.~\ref{sec:conc} by summarizing our main results and outlining future research directions in this area.

%%%%%%%%%%%%%%%%%%%%%%%%%%%%%%%%%%%%%%%%%%%%%%%%%%%%%%%%%%%%%%%%%

\section{Quark and Gluon Helicity TMDs at small $x$}
\label{sec:TMDs}

%%%%%%%%%%%%%%%%%%%%%%%%%%%%%%%%%%%%%%%%%%%%%%%%%%%%%%%%%%%%%%%%%

\subsection{Quark Helicity TMD}

We start with the quark helicity TMD defined by  \cite{Mulders:1995dh}
\begin{align} \label{e:hdef}
g_{1L}^q (x, k_T^2) = \frac{1}{(2\pi)^3} \, \half \sum_{S_L} S_L \: \int d^2 r \, dr^- \,
e^{i k \cdot r} \bra{p S_L} \bar\psi(0) \, \mathcal{U}[0,r] \, \frac{\gamma^+ \gamma^5}{2} \,
\psi(r) \ket{p S_L}_{r^+ = 0} .
\end{align}
Note that antisymmetrization over the target proton spin projection on the beam axis $S_L= \pm 1$ is ``optional''; parity symmetry guarantees that antisymmetry in the quark spin is sufficient, so that we may equivalently just set $S_L = +1$.  Our convention for light-cone coordinates is $v^\pm = (v^0 \pm v^3)/\sqrt{2}$ and the proton is moving in the light-cone ``plus" direction.

Our first goal is to simplify \eq{e:hdef} at small $x$.  The particulars of the Dirac structure in this forthcoming derivation will be specific to the helicity distribution because of the $\half \gamma^+ \gamma^5$ helicity projector in \eq{e:hdef}, but the overall approach will be common to any quark distribution at small $x$, such as transversity if we were to replace this matrix with the transversity projector $\half \gamma^5 \gamma^+ \gamma^\bot$.  

The formal definition \eqref{e:hdef} of the helicity TMD includes the process-dependent gauge link $\mathcal{U}[0,r]$; for SIDIS, the gauge link is explicitly given by
\begin{align}
\!\!\!  \mathcal{U}[0,r] = 
\mathcal{P} \exp\left[ i g \int\limits_{+\infty}^0 dz^- \, A^+ (0^+, z^- , \ul{0}) \right] 
\,
\mathcal{P} \exp\left[ - i g \int\limits_{\ul{r}}^{\ul{0}} d\ul{z} 
\cdot \ul{A} (0^+, +\infty^- , \ul{z}) \right] 
\, 
\mathcal{P} \exp\left[ i g \int\limits_{r^-}^{+\infty} dz^- \, A^+ (0^+, z^- , \ul{r}) \right],
\end{align}
where the gauge fields $A^\mu = A^{a \, \mu} t^a$ are color matrices and $t^a$ are fundamental generators of SU($N_c$).  In the $A^- = 0$ gauge that we will employ here, one can neglect the transverse link at infinity, leaving just the light-like, semi-infinite Wilson lines
\begin{align}\label{UVV}
\mathcal{U}[0,r] = V_{\ul 0} [0, \infty] \: V_{\ul r} [\infty , r^-] ,
\end{align}
where we use the following notation for the fundamental Wilson lines,
\begin{align}
  V_{\un{x}} [b^-, a^-] = \mathcal{P} \exp \left[ i g
    \int\limits_{a^-}^{b^-} d x^- \, A^+ (x^+=0, x^-, {\un x})
  \right].
\end{align}

Employing \eq{UVV} in \eq{e:hdef} we arrive at
\begin{align} \label{e:hdef2}
g_{1L}^q (x, k_T^2) = \frac{1}{(2\pi)^3} \, \int d^2 r \, dr^- \,
e^{i k \cdot r} \bra{p, \, S_L=+1} \bar\psi(0) \, V_{\ul 0} [0, \infty] \: V_{\ul r} [\infty , r^-] \, \frac{\gamma^+ \gamma^5}{2} \,
\psi(r) \ket{p, \,  S_L= +1}_{r^+ = 0} .
\end{align}
Inserting a complete set of states we get
\begin{align} \label{e:hdef3}
g_{1L}^q (x, k_T^2) = \frac{1}{(2\pi)^3} \, \sum_X \, \int d^2 r \, dr^- \,
e^{i k \cdot r} \left( \thalf \gamma^+ \gamma^5 \right)_{\alpha \beta} \, \bra{p, \, S_L = + 1} \bar\psi_\alpha (0) \, V_{\ul 0} [0, \infty] \ket{X} \notag \\ \times \: \bra{X} V_{\ul r} [\infty , r^-] \,
\psi_\beta (r) \ket{p, \, S_L = + 1}_{r^+ = 0} .
\end{align}

Using this and converting to semi-classical operator averaging used in the saturation/Color Glass Condensate (CGC) approach \cite{Gribov:1984tu,Mueller:1986wy,McLerran:1993ni,McLerran:1993ka,McLerran:1994vd,Kovchegov:1996ty,Kovchegov:1997pc,Jalilian-Marian:1997xn} (see
\cite{Gribov:1984tu,Iancu:2003xm,Weigert:2005us,Jalilian-Marian:2005jf,Gelis:2010nm,Albacete:2014fwa,KovchegovLevin}
for reviews) gives
\begin{align}\label{TMD11}
g_{1L}^q (x, k_T^2) = \frac{2 p^+}{(2\pi)^3} \: \sum_X \, \int d^{2} \zeta \, d \zeta^- \, d^{2} \xi \, d \xi^-
\, e^{i k \cdot (\zeta - \xi)} \left( \thalf \gamma^+ \gamma^5 \right)_{\alpha \beta}
\left\langle \bar\psi_\alpha (\xi) \, V_{\ul \xi} [\xi^-, \infty] \ket{X} \: \bra{X} 
 V_{\ul \zeta} [\infty , \zeta^-] \, \psi_\beta (\zeta) \right\rangle,
\end{align}
where the angle brackets denote averaging in the target shock wave \cite{Balitsky:1995ub,Balitsky:1998ya} where the target polarization $S_L = +1$ is implied, but not shown. 

\begin{figure}[ht]
\begin{center}
\includegraphics[width= 0.9 \textwidth]{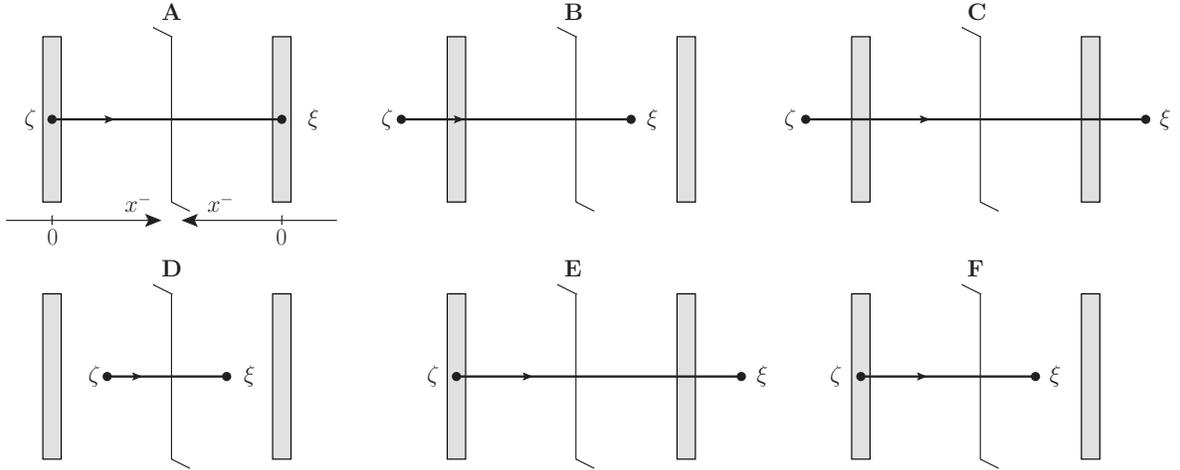} 
\caption{The main types of diagrams contributing to the quark helicity TMD in \eq{TMD11}.}
\label{fig:quark_TMD}
\end{center}
\end{figure}

Identifying the Wilson lines with the quark propagating to the final state, one can think of \eq{TMD11} as containing the inclusive quark production amplitude squared. Hence we are back to the case of SIDIS considered in \cite{Kovchegov:2015zha}. \eq{TMD11} is represented graphically in \fig{fig:quark_TMD}. There, the shaded rectangles represent the target shock wave. Thin vertical line is the final state cut. The thick horizontal lines represent the Wilson lines, which are located at different transverse plane positions $\un \zeta$ and $\un \xi$ on either side of the cut. The diagrams in \fig{fig:quark_TMD} are classified according to whether each of $\zeta^-$ and $\xi^-$ are negative, positive, or zero (corresponding to the quark field being inside the shock wave). Diagrams in the B category also include the $\zeta^- >0, \xi^- <0$ contribution, which is not shown explicitly in \fig{fig:quark_TMD}. Similarly, diagrams E also include $\zeta^- <0, \xi^- =0$ contribution, while F-graphs also include $\zeta^- > 0, \xi^- =0$ ordering: while neither of those are shown in \fig{fig:quark_TMD}, it is implied that they have to be included. 

%%%%%%%%%%%%%%%%%%%%%%%%%%%%%%%%%%%%%%%%%%%%%%%%%%%%%%%%%%%%%%%%%%%%%
%%%%%%%%%%%%%%%%%%%%%%%%%%%%%%%%%%%%%%%%%%%%%%%%%%%%%%%%%%%%%%%%%%%%%

Diagram A, evaluated at the lowest quasi-classical order with quarks exchanged in the $t$-channel interaction with the target, is the handbag diagram. Diagram B, along with the $\zeta^- >0, \xi^- <0$ contribution, is the 1-loop answer found in \cite{Kovchegov:2015zha}. Diagram D does not allow for a spin-dependent interaction with the target, since both the $\zeta$ and $\xi$ vertices are located after the shock wave. Hence, diagram D does not contribute. Diagrams C, E and F have to be investigated separately, along with the diagram A. 

\begin{figure}[ht]
\begin{center}
\includegraphics[width= 0.9 \textwidth]{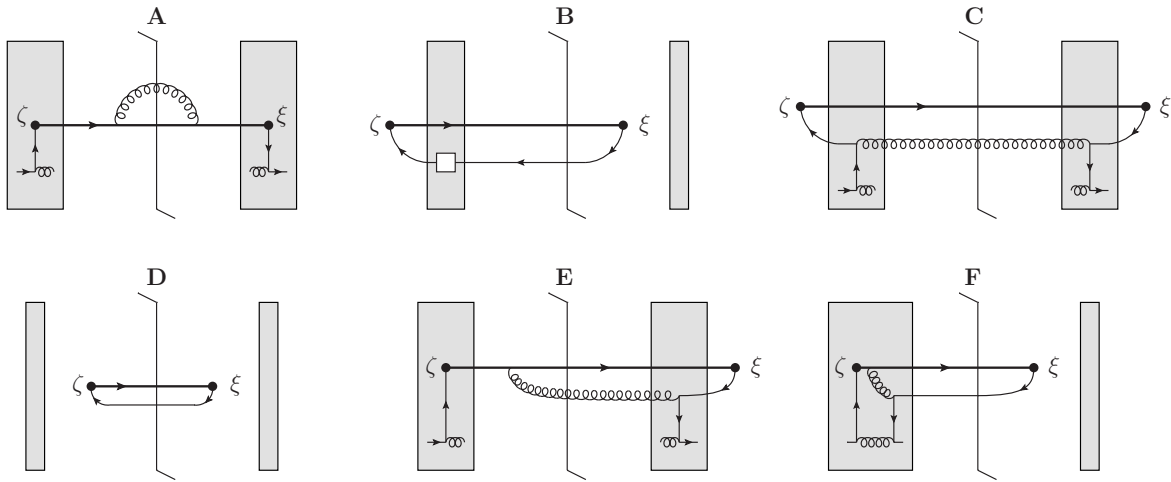} 
\caption{Diagrams contributing to the quark TMD defined in \eq{TMD11} with the order-$\as$ corrections due to $s$-channel gluon emissions shown explicitly. The thinner solid lines denote quark propagators, while the thicker solid lines are the Wilson lines (as in \fig{fig:quark_TMD}).}
\label{fig:quark_TMD_detailed}
\end{center}
\end{figure}

A more detailed representation of the types of diagrams that may contribute up to and including order-$\as$ is given in \fig{fig:quark_TMD_detailed}. (In our power counting the interactions with the shock wave are considered to be of order one.) For each diagram class we show only one sample correction: for instance, in diagram A the gluon can also be both emitted and absorbed in the amplitude or in the complex conjugate amplitude, in diagram C the $t$-channel exchanges can take place on either side of the cut, while in diagram E the gluon can be emitted from the Wilson line on either side of the cut. The box in diagram B represents spin-dependent sub-eikonal interaction with the target, following the convention introduced in \cite{Kovchegov:2015pbl}. (The interaction will be detailed below, but it includes the $t$-channel quark exchanges with the target shown in other graphs.) We will be working in $A^- =0$ light-cone gauge throughout this paper. 

Diagram C appears to contribute: however, the interactions of the quark like with the target cancel if we move the $t$-channel exchanges across the cut \cite{Kovchegov:2015zha}. Hence diagram C does not contribute at this order. (At the order-$\as^2$ diagram C can contribute: this is the order beyond the one considered explicitly here. Still we believe that the leading-logarithmic contribution of diagram C will be canceled even at that order in the coupling due to the same mechanism as described in Appendix~\ref{A}.) Diagram F is energy suppressed, since the gluon in it has to be emitted and absorbed over a very short lifetime of the shock wave in the $x^-$ direction. (Moreover, at small $x$ the gluon in diagram F has to carry the same large ``minus" momentum as the $s$-channel antiquark propagator connecting to the vertex $\xi$: the merger of this gluon with the Wilson line that begins at $\zeta$ cannot be eikonal, since the gluon and the quark propagator that this Wilson line represents carry comparable ``minus" momenta.) Diagram E may contribute (as we have mentioned, the gluon there may connect to either one of the Wilson lines to the left and right of the cut). Diagram A may be ``dressed" by gluon interactions with the Wilson lines, as shown in \fig{fig:quark_TMD_detailed}. In Appendix~\ref{A} we show that the diagrams A and E cancel at the order-$\as$ considered in \fig{fig:quark_TMD_detailed} (and in the leading logarithmic approximation in $x$) and discuss what happens at higher orders in the coupling.\footnote{Unlike the other diagrams considered here, the diagram A does appear to contribute at order-1, that is, at Born level, when no gluon emission corrections like those shown in \fig{fig:quark_TMD_detailed} are included. However, such contribution is independent of $x$, and is subleading compared to the contribution of diagram B we calculate below, which grows as a power of $1/x$.}

%%%%%%%%%%%%%%%%%%%%%%%%%%%%%%%%%%%%%%%%%%%%%%%%%%%%%%%%%%%%%%%%%%%%%
%%%%%%%%%%%%%%%%%%%%%%%%%%%%%%%%%%%%%%%%%%%%%%%%%%%%%%%%%%%%%%%%%%%%%

\begin{figure}[ht]
\begin{center}
\includegraphics[width= 0.45 \textwidth]{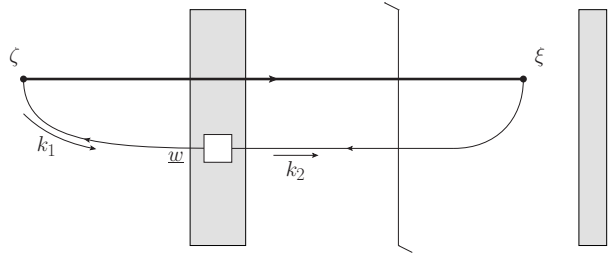} 
\caption{A more detailed illustration of diagram B. }
\label{fig:B}
\end{center}
\end{figure}

We conclude that the diagrams A and C-F do not contribute at the leading small-$x$ level. Therefore, we are left with the diagram B and its ``mirror image", the  contribution with $\zeta^- >0, \xi^- <0$. The ``mirror image" is just  complex conjugate of the diagram B. Hence, diagram B and its ``mirror image" give
\begin{align}\label{TMD12}
g_{1L}^q (x, k_T^2) = \frac{2 p^+}{(2\pi)^3} \,  \sum_{\bar{q}} \, \int\limits_{-\infty}^0 d \zeta^- \, \int\limits^{\infty}_0 d \xi^- \, \int d^{2} \zeta \, d^{2} \xi 
\, e^{i k \cdot (\zeta - \xi)} \left( \thalf \gamma^+ \gamma^5 \right)_{\alpha \beta}
 \, & \left\langle  \bar\psi_\alpha (\xi)  \, V_{\ul \xi} [\xi^-, \infty] \ket{\bar{q}} \: \bra{\bar{q}}
 V_{\ul \zeta} [\infty , \zeta^-] \,  \psi_\beta (\zeta) \right\rangle \notag \\ + & \mbox{c.c.} ,
\end{align}
where we have replaced $X \to \bar{q}$ since only the antiquark is produced in the final state (in addition to the quark represented by the Wilson lines). The sum $\sum_{\bar{q}}$ now denotes the Lorentz-invariant integral over the antiquark momentum and a sum over its polarizations and colors. Putting $V_{\ul \xi} [\xi^-, \infty]  =1$ for $\xi^- >0$ (since the Wilson line does not cross the shock wave, it is trivial) and replacing $V_{\ul \zeta} [\infty , \zeta^-] \to V_{\ul \zeta} [\infty , -\infty]$ for $\zeta^- <0$ since this Wilson line crosses the shock wave and gets all the non-trivial contributions from this crossing only, we simplify \eqref{TMD12} to
\begin{align}\label{TMD13}
g_{1L}^q (x, k_T^2) = \frac{2 p^+}{(2\pi)^3} \,  \sum_{\bar{q}} \, \int\limits_{-\infty}^0 d \zeta^- \, \int\limits^{\infty}_0 d \xi^- \, \int d^{2} \zeta \, d^{2} \xi 
\, e^{i k \cdot (\zeta - \xi)} \left( \thalf \gamma^+ \gamma^5 \right)_{\alpha \beta}
 \, \left\langle  \bar\psi_\alpha (\xi)  \ket{\bar{q}} \: \bra{\bar{q}} \,
 V_{\ul \zeta} [\infty , -\infty] \, \psi_\beta (\zeta) \right\rangle + \mbox{c.c.} . 
\end{align}
Diagram B is illustrated in a little more detail in \fig{fig:B}.

%%%%%%%%%%%%%%%%%%%%%%%%%%%%%%%%%%%%%%%%%%%%%%%%%%%%%%%%%%%%%%%%%%%%%
%%%%%%%%%%%%%%%%%%%%%%%%%%%%%%%%%%%%%%%%%%%%%%%%%%%%%%%%%%%%%%%%%%%%%

In evaluating the diagram B we impose the $\zeta^- <0, \xi^- >0$ ordering, which makes sure that the the vertices at $\zeta$ and $\xi$ are outside the shock wave. For the antiquark (background-field  \cite{Balitsky:1995ub,Balitsky:1998ya}) propagator in \fig{fig:B} traversing the shock wave we write
\begin{align}\label{propagator1}
\contraction
{}
{\bar\psi^i_\alpha}
{(\eta) \:}
{\psi^j_\beta}
\bar\psi^i_\alpha (\xi) \: \psi^j_\beta (\zeta)
&= \int d^2 w \, \frac{d^4 k_1}{(2\pi)^4} \frac{d^4 k_2}{(2\pi)^4} \, e^{i k_1^+ \zeta^-} \, 
e^{i \ul{k}_1 \cdot (\ul{w} - \ul{\zeta})} \, e^{-i k_2^+ \xi^-} \, e^{i \ul{k}_2 \cdot (\ul{\xi} - \ul{w})}
\notag \\ & \hspace{2cm} \times
\left\{ \left[ \frac{- i \slashed{k_1}}{k_1^2 + i\epsilon} \right]
\left[ \left( \hat{V}_{{\un w}}^\dagger \right)^{ji} \, (2\pi) \, \delta(k_1^- - k_2^-) \right]
\left[ - \slashed{k_2} \, (2 \pi) \, \delta (k_2^2) \right] \right\}_{\beta \alpha} ,
\end{align}
where $i$ and $j$ are the quark color indices. This quantity is constructed as an antiquark propagator: created by ${\hat d}^\dagger$ in $\psi_\beta (\zeta)$ and propagating with positive energy $k_1^-$ through the shockwave to be annihilated by $\hat d$ in $\bar\psi_\alpha(\xi)$.  Note that the ``vertex" for the antiquark passing through the shockwave (the box in \fig{fig:B}) is a Dirac matrix, denoted by $(\hat{V}^\dagger)$. Its exact structure will be clarified later, but for now, in our small-$x$ approximation, we can think of it as a light-cone Wilson line with or without an insertion of a non-eikonal local operator. 

In order to simplify the propagator \eqref{propagator1} we first integrate over $k_2^-$,  and over $k_1^+$ with $k_2^+$, while keeping in mind that $\zeta^- <0$ and $\xi^- >0$.\footnote{Here and below in this paper, when calculating diagrams like B, we will neglect the non-logarithmic instantaneous terms; in this case the instantaneous term leads to the delta-function $\delta (\zeta^-)$ confining the corresponding vertex to the inside of the shockwave: such terms contribute to diagrams A, E and F only. This means that in \eq{propagator1} one should understand $\slashed{k_1}$ as $\gamma^- \frac{{\un k}_1^2}{2 k_1^-} + \gamma^+ k_1^- - {\un \gamma} \cdot {\un k}_1$.} This yields
\begin{align}\label{propagator2}
\contraction
{}
{\bar\psi^i_\alpha}
{(\eta) \:}
{\psi^j_\beta}
\bar\psi^i_\alpha (\xi) \: \psi^j_\beta (\zeta)
&= \int  d^2 w \, \frac{d^2 k_1 \, d k_1^-}{(2\pi)^3} \, \frac{d^2 k_2}{(2\pi)^2} \, e^{i \frac{{\un k}_1^2}{2 k_1^-} \zeta^- - i \frac{{\un k}_2^2}{2 k_1^-} \xi^- + i \ul{k}_1 \cdot (\ul{w} - \ul{\zeta}) + i \ul{k}_2 \cdot (\ul{\xi} - \ul{w})} \, \theta (k_1^-)
\notag \\ & \hspace{2cm} \times
\left\{ \left[ \frac{\slashed{k_1}}{2 k_1^-} \right]
\left[ \left( \hat{V}_{{\un w}}^\dagger \right)^{ji} \right]
\left[ \frac{ \slashed{k_2}}{2 k_1^-} \right] \right\}_{\beta \alpha} \Bigg|_{k_2^- = k_1^-, k_1^2 =0, k_2^2 =0} .
\end{align}
Plugging this back into \eq{TMD13} and integrating over $\un \xi$  and ${\un k}_2$ we obtain
\begin{align}\label{TMD14}
g_{1L}^q (x, k_T^2) & = \frac{2 p^+}{(2\pi)^3}  \: \int\limits_{-\infty}^0 d \zeta^- \, \int\limits^{\infty}_0 d \xi^- \, \int d^{2} \zeta 
\, e^{i k^+ (\zeta^- - \xi^-)} \left( \thalf \gamma^+ \gamma^5 \right)_{\alpha \beta}
 \, \Bigg\langle \mbox{T} \; V_{\ul \zeta}^{ij} [\infty , -\infty] \,  \int  d^2 w \, \frac{d^2 k_1 \, d k_1^-}{(2\pi)^3}   \\ & \times e^{i \frac{{\un k}_1^2}{2 k_1^-} \zeta^- - i \frac{{\un k}^2}{2 k_1^-} \xi^- + i (\ul{k}_1 + {\un k}) \cdot (\ul{w} - \ul{\zeta})} \, \theta (k_1^-) 
\left\{ \left[ \frac{\slashed{k_1}}{2 k_1^-} \right]
\left[ \left( \hat{V}_{{\un w}}^\dagger \right)^{ji} \right]
\left[ \frac{ \slashed{k_2}}{2 k_1^-} \right] \right\}_{\beta \alpha} \Bigg\rangle \Bigg|_{k_2^- = k_1^-, k_1^2 =0, k_2^2 =0, {\un k}_2 = - {\un k}}  + \mbox{c.c.} . \notag
\end{align}
Here we explicitly insert the time-ordering sign T, which is often omitted but implied in the CGC calculations. Since both ``Wilson lines" $V_{\ul \zeta}$ and $\hat{V}_{{\un w}}$ are in the amplitude of diagram B in \fig{fig:B}, the come in with a time-ordering sign. Distinguishing time-ordered and anti-time ordered correlation functions will be very important below. As we detail in Appendix~\ref{B}, inserting time-ordering sign T and the anti-time ordering sign $\overline{\mbox{T}}$ allow us to distinguish amplitudes from the complex conjugate amplitudes. This way we are able to differentiate between an expectation value of the complex conjugate operator in the amplitude versus the complex conjugate of the expectation value of the operator in the amplitude.

Next we integrate over $\zeta^-$ and $\xi^-$. This yields
\begin{align}\label{TMD25}
g_{1L}^q (x, k_T^2) & = - \frac{2 p^+}{(2\pi)^3}  \: \int d^{2} \zeta 
\, d^2 w \, \frac{d^2 k_1 \, d k_1^-}{(2\pi)^3} \, e^{i (\ul{k}_1 + {\un k}) \cdot (\ul{w} - \ul{\zeta})} \, \theta (k_1^-) \, \Bigg\langle \mbox{T} \, V_{\ul \zeta}^{ij} [\infty , -\infty]   
\, \Tr \left[ \thalf \gamma^+ \gamma^5 \, \slashed{k_1} \,
\left( \hat{V}_{{\un w}}^\dagger \right)^{ji}  \,
\slashed{k_2} \right]  \Bigg\rangle \\ & \times  \, \frac{1}{\left[ 2 k_1^- k^+ + {\un k}_1^2 - i \epsilon k_1^-  \right] \, \left[ 2 k_1^- k^+ + {\un k}^2 + i \epsilon k_1^- \right]}  \Bigg|_{k_2^- = k_1^-, k_1^2 =0, k_2^2 =0, {\un k}_2 = - {\un k}}  + \mbox{c.c.} . \notag
\end{align}

Introducing polarization sums, we write
\begin{align}\label{TMD15}
g_{1L}^q (x, k_T^2) & = - \frac{2 p^+}{(2\pi)^3}  \:  \int d^{2} \zeta 
\, d^2 w \, \frac{d^2 k_1 \, d k_1^-}{(2\pi)^3} \, e^{i (\ul{k}_1 + {\un k}) \cdot (\ul{w} - \ul{\zeta})} \, \theta (k_1^-) \, \sum_{\sigma_1, \, \sigma_2} {\bar v}_{\sigma_2} (k_2)  \thalf \gamma^+ \gamma^5 v_{\sigma_1} (k_1)  \, \Bigg\langle \mbox{T} \, V_{\ul \zeta}^{ij} [\infty , -\infty]   \\ & \times 
\ {\bar v}_{\sigma_1} (k_1) \left( \hat{V}_{{\un w}}^\dagger \right)^{ji} v_{\sigma_2} (k_2)  \Bigg\rangle \, \frac{1}{\left[ 2 k_1^- k^+ + {\un k}_1^2 - i \epsilon k_1^-  \right] \, \left[ 2 k_1^- k^+ + {\un k}^2 + i \epsilon k_1^- \right]}  \Bigg|_{k_2^- = k_1^-, k_1^2 =0, k_2^2 =0, {\un k}_2 = - {\un k}}  + \mbox{c.c}. \notag
\end{align}

Further, we define the (antiquark) polarized ``Wilson line" as a longitudinal spin-dependent part of an antiquark scattering amplitude on the shock wave. We need the part of the scattering amplitude proportional to the Pauli matrix $\sigma^3$ in helicity space, that is, to $\sigma \, \delta_{\sigma \sigma'}$ with $\sigma$ and $\sigma'$ the helicities of the antiquark before and after scattering respectively. Using the Brodsky-Lepage (BL) spinors \cite{Lepage:1980fj} we write
\begin{align}\label{spinors1}
\left[ {\bar v}_{\sigma} (p) \, \Big( \hat{V}^\dagger_{\ul x} \Big) \, v_{\sigma'} (p') \right] &=
2 \sqrt{p^- p^{\prime \, -}} \delta_{\sigma \sigma'} \left( V^\dagger_{\ul x} - \sigma V_{\ul x}^{pol \, \dagger} + \ldots \right) 
= 2 \sqrt{p^- p^{\prime \, -}} \delta_{\sigma \sigma'} \: V^\dagger_{\ul x} (-\sigma) + \ldots ,
\end{align}
where the ellipsis denote the sub-eikonal corrections independent of helicity, which we are not interested in and we use a shorthand notation $V_{\un x} \equiv V_{\un x} [\infty, -\infty]$.\footnote{The general convention for BL spinors is as follows (these matrix elements appear either in the scattering amplitude or in the complex conjugate amplitude):
\begin{subequations}
\begin{align}
& {\bar u}_{\sigma} (p) \, \Big( \hat{V}_{\ul x} \Big) \, u_{\sigma'} (p') = 2 \sqrt{p^- p^{\prime \, -}} \delta_{\sigma \sigma'} \: V_{\ul x} (\sigma) , \ \ \ {\bar u}_{\sigma} (p) \, \Big( \hat{V}^\dagger_{\ul x} \Big) \, u_{\sigma'} (p') = 2 \sqrt{p^- p^{\prime \, -}} \delta_{\sigma \sigma'} \: V^\dagger_{\ul x} (\sigma) , \\ & {\bar v}_{\sigma} (p) \, \Big( \hat{V}_{\ul x} \Big) \, v_{\sigma'} (p') = 2 \sqrt{p^- p^{\prime \, -}} \delta_{\sigma \sigma'} \: V_{\ul x} (-\sigma) , \ \ \ {\bar v}_{\sigma} (p) \, \Big( \hat{V}^\dagger_{\ul x} \Big) \, v_{\sigma'} (p') = 2 \sqrt{p^- p^{\prime \, -}} \delta_{\sigma \sigma'} \: V^\dagger_{\ul x} (-\sigma) .
\end{align}
\end{subequations}
}

In addition, we will employ
\begin{align}\label{spinors2}
{\bar v}_{\sigma_2} (k_2) \thalf \gamma^+ \gamma^5 v_{\sigma_1} (k_1)  = 
\half \sigma_2 \, \delta_{\sigma_2 \sigma_1} 
\, \frac{(\ul{k}_2 \cdot \ul{k}_1) - i \sigma_2 (\ul{k}_2 \times \ul{k}_1)}
{\sqrt{k_1^- k_2^-}}  
\end{align}
for the $(\pm)$-interchanged Brodsky-Lepage spinors (which we will also refer to as the anti-BL spinors).

Using the matrix elements from Eqs.~\eqref{spinors1} and \eqref{spinors2} in \eq{TMD15} we obtain
\begin{align}\label{TMD16}
g_{1L}^q (x, k_T^2) & =  \frac{2 p^+}{(2\pi)^3}  \: \int d^{2} \zeta 
\, d^2 w \, \frac{d^2 k_1 \, d k_1^-}{(2\pi)^3} \, e^{i (\ul{k}_1 + {\un k}) \cdot (\ul{w} - \ul{\zeta})} \, \theta (k_1^-) \, \sum_{\sigma_1} \, \sigma_1 \, \left[ {\un k} \cdot {\un k}_1 - i \sigma_1 \, {\un k} \times {\un k}_1 \right]  \notag \\ & \times \, \left\langle \mbox{T} \, \mbox{tr} \left[ V_{\ul \zeta}^{ij} [\infty , -\infty] \, V_{{\un w}}^\dagger (-\sigma_1) \right]  \right\rangle 
\ \frac{1}{\left[ 2 k_1^- k^+ + {\un k}_1^2 - i \epsilon k_1^-  \right] \, \left[ 2 k_1^- k^+ + {\un k}^2 + i \epsilon k_1^- \right]} + \mbox{c.c.} .
\end{align}
Remembering that $k^+ = x \, p^+$ and we are considering the small-$x$ regime (and, hence, $2 k_1^- k^+ = 2 k_1^- x p^+ \ll {\un k}^2, {\un k}_1^2$), we get
\begin{align}\label{TMD17}
g_{1L}^q (x, k_T^2) = & \, \frac{2 p^+}{(2\pi)^3}  \:  \int d^{2} \zeta 
\, d^2 w \, \frac{d^2 k_1 \, d k_1^-}{(2\pi)^3} \, e^{i (\ul{k}_1 + {\un k}) \cdot (\ul{w} - \ul{\zeta})} \, \theta (k_1^-) \notag \\ & \times \, \sum_{\sigma_1} \, \sigma_1 \, \frac{{\un k} \cdot {\un k}_1 - i \sigma_1 \, {\un k} \times {\un k}_1}{{\un k}_1^2  \, {\un k}^2 }  \, \left\langle \mbox{T} \, \mbox{tr} \left[ V_{\ul \zeta} \, V_{{\un w}}^\dagger (- \sigma_1) \right] \right\rangle + \mbox{c.c.} .
\end{align}
Writing $V_{{\un w}}^\dagger (-\sigma_1) = V_{{\un w}}^\dagger - \sigma_1 \, V_{{\un w}}^{pol \, \dagger}$ allows us to sum over $\sigma_1$ obtaining, after performing ${\un k}_1$ integration as well, 
\begin{align}\label{TMD18}
g_{1L}^q (x, k_T^2) & =  \frac{4 p^+ i}{(2\pi)^4}  \: \int d^{2} \zeta 
\, d^2 w \, e^{ - i {\un k} \cdot (\ul{\zeta} - \ul{w})}  \, \int\limits_0^\infty  \frac{d k_1^-}{2\pi}  \\ & \times  \, \left\{  \frac{\un{k} }{\un{k}^2} \cdot \frac{\un{\zeta} - \un{w}}{|\un{\zeta} - \un{w}|^2}   \, \left\langle \mbox{T} \, \mbox{tr} \left[ V_{\ul \zeta} \, V_{{\un w}}^{pol \, \dagger} \right] + \bar{\mbox{T}} \, \mbox{tr} \left[ V_{\un \zeta}^{pol} \, V_{\ul w}^\dagger \right] \right\rangle + i \, \frac{\un{k} }{\un{k}^2} \times \frac{\un{\zeta} - \un{w}}{|\un{\zeta} - \un{w}|^2}  \, \left\langle \mbox{T} \, \mbox{tr} \left[ V_{\ul \zeta} \, V_{{\un w}}^\dagger \right] - \bar{\mbox{T}} \, \mbox{tr} \left[ V_{\ul \zeta} \, V_{{\un w}}^\dagger \right]  \right\rangle \right\}, \notag
\end{align}
where we have explicitly added the complex conjugate term (and interchanged ${\un \zeta} \leftrightarrow {\un w}$ in it) by employing the fact that $[\mbox{T} O_1 (x) \, O_2 (y)]^\dagger = \bar{\mbox{T}} O_2^\dagger (y) \, O_1^\dagger (x)$ for two operators $O_1 (x)$ and $O_2 (y)$. As mentioned before, the sign $\bar{\mbox{T}}$ denotes anti-time ordering. 

Before we continue, let us stress the importance of the ordering of (polarized and/or unpolarized) Wilson lines in the non time-ordered correlation functions. To do this, let us introduce the following useful relations between the expectation values of Wilson lines:
\begin{subequations}\label{Wilson_rels}
\begin{align}
\left\langle \mbox{T} \, \mbox{tr} \left[ V_{\ul x} \, V_{{\un y}}^{pol \, \dagger} \right]  \right\rangle = \left\langle \mbox{tr} \left[ V_{\ul x} \, V_{{\un y}}^{pol \, \dagger} \right]  \right\rangle ,  \label{Wilson_rels_a} \\
\left\langle \bar{\mbox{T}} \, \mbox{tr} \left[ V_{\ul x} \, V_{{\un y}}^{pol \, \dagger} \right]  \right\rangle = \left\langle \mbox{tr} \left[ V_{{\un y}}^{pol \, \dagger} \, V_{\ul x} \right]  \right\rangle . \label{Wilson_rels_b}
\end{align}
\end{subequations}
The relations Eqs.~\eqref{Wilson_rels} are written here for one regular Wilson line and for one unpolarized Wilson line: however, they are also valid for correlators of two regular Wilson lines. The order of the Wilson lines under the trace matters for the right-hand sides of Eqs.~\eqref{Wilson_rels}. In the sense of \eq{TMD11}, the right Wilson line in each non-time-ordered correlator can be thought of as contributing to the amplitude, while the left Wilson line contributes to the complex conjugate amplitude. The Wilson lines are bosonic operators (even the polarized ``Wilson lines" are bosonic, as we will see below): hence the ordering of the Wilson lines is not important for the (anti-)time-ordered correlation functions. 

Note that this ordering issue does not apply to the standard eikonal CGC calculations done in the leading-logarithmic approximation (LLA)
\cite{Jalilian-Marian:1997dw,Jalilian-Marian:1997gr,Iancu:2001ad,Iancu:2000hn,Balitsky:1995ub,Balitsky:1998ya,Kovchegov:1999yj,Kovchegov:1999ua}, where all the Wilson lines are standard eikonal Wilson lines, and the background gluon field is assumed to be classical \cite{McLerran:1993ni,McLerran:1993ka,McLerran:1994vd,Kovchegov:1996ty,Kovchegov:1997pc,Jalilian-Marian:1997xn} rather than being an operator: in such case the order of Wilson lines does not matter and Eqs.~\eqref{Wilson_rels} are trivially satisfied (see \cite{Kovchegov:2001sc} for applications of that result to inclusive gluon production). The relations \eqref{Wilson_rels} were shown to work in \cite{Mueller:2012bn} up to next-to-leading logarithms (NLL) in $x$ for the unpolarized BK/JIMWLK evolution. 

\begin{figure}[ht]
\begin{center}
\includegraphics[width= 0.65 \textwidth]{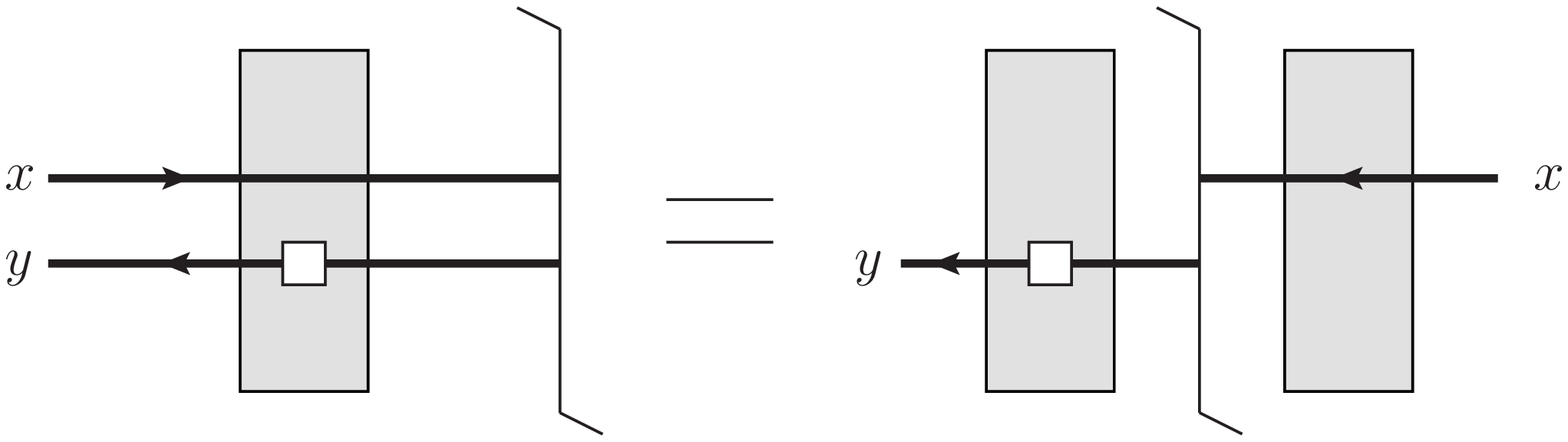} 
\caption{The reflection of the Wilson line from the complex conjugate amplitude to the amplitude discussed in the text.}
\label{fig:flip}
\end{center}
\end{figure}

Diagrammatically the relations \eqref{Wilson_rels} can be pictured as arising from the reflection symmetry of light-cone Wilson lines (true ``unpolarized" Wilson lines) with respect to the final state cut. \eq{Wilson_rels_a} can be thought of as being due to reflecting a light-cone Wilson line from the complex conjugate amplitude (on the expressions right-hand side) back into the amplitude (the left-hand side of \eq{Wilson_rels_a}), as illustrated in \fig{fig:flip}. \eq{Wilson_rels_b} arises after the reflection of the light-cone Wilson line from the amplitude into the complex conjugate amplitude. 

%%%%%%%%%%%%%%%%%%%%%%%%%%%%%%%%%%%%%%%%%%%%%%%%%%%%%%%%%%%%%%%%%%%%%
%%%%%%%%%%%%%%%%%%%%%%%%%%%%%%%%%%%%%%%%%%%%%%%%%%%%%%%%%%%%%%%%%%%%

Taking the real and imaginary parts of Eqs.~\eqref{Wilson_rels} one obtains more useful formulas
\begin{subequations}\label{Wilson_rels2}
\begin{align}
2 \: \mbox{Re} \,  \left\langle \mbox{T} \, \mbox{tr} \left[ V_{\ul x} \, V_{{\un y}}^{pol \, \dagger} \right]  \right\rangle = \left\langle \mbox{tr} \left[ V_{\ul x} \, V_{{\un y}}^{pol \, \dagger} \right]  \right\rangle + \left\langle \mbox{tr} \left[ V_{{\un y}}^{pol} \, V_{\ul x}^ \dagger \right]  \right\rangle, \\
2 \: i \, \mbox{Im} \,  \left\langle \mbox{T} \, \mbox{tr} \left[ V_{\ul x} \, V_{{\un y}}^{pol \, \dagger} \right]  \right\rangle = \left\langle \mbox{tr} \left[ V_{\ul x} \, V_{{\un y}}^{pol \, \dagger} \right]  \right\rangle - \left\langle \mbox{tr} \left[ V_{{\un y}}^{pol} \, V_{\ul x}^ \dagger \right]  \right\rangle , \\
2 \: \mbox{Re} \,  \left\langle \bar{\mbox{T}} \, \mbox{tr} \left[ V_{\ul x} \, V_{{\un y}}^{pol \, \dagger} \right]  \right\rangle = \left\langle \mbox{tr} \left[ V_{{\un y}}^{pol \, \dagger} \, V_{\ul x} \right]  \right\rangle + \left\langle \mbox{tr} \left[ V_{\ul x}^ \dagger  \, V_{{\un y}}^{pol} \right]  \right\rangle , \\
2 \: i \, \mbox{Im} \,  \left\langle \bar{\mbox{T}} \, \mbox{tr} \left[ V_{\ul x} \, V_{{\un y}}^{pol \, \dagger} \right]  \right\rangle = \left\langle \mbox{tr} \left[ V_{{\un y}}^{pol \, \dagger} \, V_{\ul x}  \right]  \right\rangle - \left\langle \mbox{tr} \left[ V_{\ul x}^ \dagger \, V_{{\un y}}^{pol} \right]  \right\rangle .
\end{align}
\end{subequations}

Returning to \eq{TMD18} we notice that 
\begin{align}
\left\langle \mbox{T} \, \mbox{tr} \left[ V_{\ul \zeta} \, V_{{\un w}}^\dagger \right] - \bar{\mbox{T}} \, \mbox{tr} \left[ V_{\ul \zeta} \, V_{{\un w}}^\dagger \right]  \right\rangle = \left\langle  \mbox{tr} \left[ V_{\ul \zeta} \, V_{{\un w}}^\dagger \right] - \mbox{tr} \left[ V_{{\un w}}^\dagger \, V_{\ul \zeta} \right]  \right\rangle = 0,
\end{align}
since for true Wilson lines the reflection symmetries that led to Eqs.~\eqref{Wilson_rels} also imply that $\left\langle  \mbox{tr} \left[ V_{\ul \zeta} \, V_{{\un w}}^\dagger \right] \right\rangle = \left\langle   \mbox{tr} \left[ V_{{\un w}}^\dagger \, V_{\ul \zeta} \right]  \right\rangle$ (with the same NLL accuracy as Eqs.~\eqref{Wilson_rels} were verified up to). (Note that $\mbox{T} \, \mbox{tr} \left[ V_{\ul \zeta} \, V_{{\un w}}^\dagger \right] - \bar{\mbox{T}} \, \mbox{tr} \left[ V_{\ul \zeta} \, V_{{\un w}}^\dagger \right]$ is not the odderon operator. The latter is $\mbox{tr} \left[ V_{\ul \zeta} \, V_{{\un w}}^\dagger \right]  - \mbox{tr} \left[ V_{{\un w}}  \, V_{\ul \zeta}^\dagger  \right]$ \cite{Kovchegov:2003dm,Hatta:2005as} and it gives zero after the impact parameter integration, as observed in \cite{Kovchegov:2012ga}.)

Since the second term in the curly brackets of \eq{TMD18} is zero, we arrive at
\begin{align}\label{TMD19}
g_{1L}^q (x, k_T^2) & =  \frac{4 p^+ i}{(2\pi)^4} \:  \int d^{2} \zeta 
\, d^2 w \, e^{ - i {\un k} \cdot (\ul{\zeta} - \ul{w})} \, \int\limits_0^\infty \frac{d k_1^-}{2\pi} \, \frac{\un{k} }{\un{k}^2} \cdot \frac{\un{\zeta} - \un{w}}{|\un{\zeta} - \un{w}|^2}  \, \left\langle  \mbox{T} \,  \mbox{tr} \left[ V_{\ul \zeta} \, V_{{\un w}}^{pol \, \dagger} \right] + \bar{\mbox{T}} \, \mbox{tr} \left[ V_{\un \zeta}^{pol} \, V_{\ul w}^\dagger \right]  \right\rangle .
\end{align}

In the flavor-singlet case that we are primarily interested in here one adds the anti-quark TMD contribution. This yields
\begin{align}\label{TMD21}
& g_{1L}^S (x, k_T^2) = \frac{4 p^+ i}{(2\pi)^4} \:  \int d^{2} \zeta 
\, d^2 w \, e^{ - i {\un k} \cdot (\ul{\zeta} - \ul{w})} \, \int\limits_0^\infty \frac{d k_1^-}{2\pi} \,  \frac{\un{k} }{\un{k}^2} \cdot  \frac{\un{\zeta} - \un{w}}{|\un{\zeta} - \un{w}|^2}  \\ & \times \,  \left\langle \mbox{T} \, \mbox{tr} \left[ V_{\ul \zeta} \, V_{{\un w}}^{pol \, \dagger} \right] + \mbox{T} \, \mbox{tr} \left[ V_{{\un w}}^{pol} \, V_{\ul \zeta}^\dagger \right] + \bar{\mbox{T}} \, \mbox{tr} \left[ V_{\un \zeta}^{pol} \, V_{\ul w}^\dagger \right]  + \bar{\mbox{T}} \, \mbox{tr} \left[ V_{\ul w} \, V_{\un \zeta}^{pol \, \dagger} \right]  \right\rangle . \notag 
\end{align}
Here we have used (again for the anti-BL spinors)
\begin{align}\label{spinors22}
{\bar u}_{\sigma_2} (k_2) \thalf \gamma^+ \gamma^5 u_{\sigma_1} (k_1)  = 
- \half \sigma_2 \, \delta_{\sigma_2 \sigma_1} 
\, \frac{(\ul{k}_2 \cdot \ul{k}_1) + i \sigma_2 (\ul{k}_2 \times \ul{k}_1)}
{\sqrt{k_1^- k_2^-}}  .
\end{align}

We next define the (flavor-singlet) polarized dipole amplitude 
\begin{align}\label{Gdefwz}
G_{{\un w}, {\un \zeta}} (zs) = \frac{k_1^- \, p^+}{N_c}  \:  \mbox{Re} \:  \left\langle \mbox{T} \, \mbox{tr} \left[ V_{\ul \zeta} \, V_{{\un w}}^{pol \, \dagger} \right] + \mbox{T} \, \mbox{tr} \left[ V_{{\un w}}^{pol} \, V_{\ul \zeta}^\dagger \right] \right\rangle 
\end{align}
with $z s = 2 k_1^- \, p^+$. This definition is different from the one used in our previous works \cite{Kovchegov:2015pbl,Kovchegov:2016weo,Kovchegov:2016zex,Kovchegov:2017jxc,Kovchegov:2017lsr} by the real-part operator Re and by the time-ordering signs shown explicitly here while they were only implied in our earlier works, as is customary in the saturation/CGC calculations (with the exception of \cite{Mueller:2012bn}). All the calculations performed in \cite{Kovchegov:2015pbl,Kovchegov:2016weo,Kovchegov:2016zex,Kovchegov:2017jxc,Kovchegov:2017lsr} were not affected by the omitted time-ordering signs, since they were {\sl de facto} applied. Similarly, the Re sign was {\sl de facto} applied as well, since only cut diagrams were calculated. In DLA, the real-part operator only makes a difference when evaluating the initial conditions for the (linear) small-$x$ evolution of the polarized dipole amplitude \eqref{Gdefwz}. In calculating these initial conditions the Re operator from the right of \eq{Gdefwz} was applied: we calculated the scattering cross sections for the Born-level processes, instead of the whole forward amplitude. (That is, we calculated the imaginary part of the forward scattering amplitude.) In Appendix~\ref{B} we show explicitly how the calculations carried out earlier in Sec.~II A of \cite{Kovchegov:2016zex} are equivalent to \eq{Gdefwz}, thus also illustrating how Eqs.~\eqref{Wilson_rels} work.

Employing the definition \eqref{Gdefwz} in \eq{TMD21} along with Eqs.~\eqref{Wilson_rels} and their complex conjugates (or, equivalently, Eqs.~\eqref{Wilson_rels2} and their complex conjugates) while noticing that for a longitudinally polarized target, due to the absence of any preferred transverse direction,
\begin{align}\label{bint2}
\int d^2 \left( \frac{\zeta + w}{2} \right) \, \left\langle \mbox{T} \, \mbox{tr} \left[ V_{\ul \zeta} \, V_{{\un w}}^{\dagger} \right] \right\rangle  \equiv h(|{\un \zeta} - {\un w}|) = h(|{\un w} - {\un \zeta}|) = \int d^2 \left( \frac{\zeta + w}{2} \right) \, \left\langle \mbox{T} \, \mbox{tr} \left[ V_{\ul w} \, V_{{\un \zeta}}^{\dagger} \right] \right\rangle
\end{align}
for correlators made out of both polarized and unpolarized Wilson lines with time-ordering and anti-time ordering, we arrive at
\begin{align}\label{TMD21.5}
g_{1L}^S (x, k_T^2) = \frac{8 \, N_c \, i}{(2\pi)^5} \: \int d^{2} \zeta 
\, d^2 w \, e^{ - i {\un k} \cdot (\ul{\zeta} - \ul{w})} \, \int\limits_{\Lambda^2/s}^1 \frac{d z}{z}   \, \frac{\un{\zeta} - \un{w}}{|\un{\zeta} - \un{w}|^2} \cdot  \frac{\un{k} }{\un{k}^2} \, G_{{\un w}, {\un \zeta}} (zs)  . 
\end{align}
Here $s \approx Q^2/x$ is the center-of-mass energy squared, while $\Lambda$ is an infrared (IR) cutoff with $\Lambda^2/s$ the lowest possible value of the variable $z$. Introducing a dummy transverse vector variable $\un y$ we rewrite \eq{TMD21.5} as
\begin{align}\label{TMD22}
g_{1L}^S (x, k_T^2) =  \frac{8 N_c}{(2\pi)^6}  \: \int d^{2} \zeta 
\, d^2 w \, d^2 y \, e^{ - i {\un k} \cdot (\ul{\zeta} - \ul{y})} \, \int\limits_{\Lambda^2/s}^1 \frac{d z}{z}  \,\frac{\un{\zeta} - \un{w}}{|\un{\zeta} - \un{w}|^2} \cdot  \frac{\un{y} - \un{w}}{|\un{y} - \un{w}|^2} \, G_{{\un w}, {\un \zeta}} (zs) ,
\end{align}
in complete agreement with Eq. (8c) in \cite{Kovchegov:2016zex}, or, equivalently, Eq. (15) in \cite{Kovchegov:2015pbl}. 

The corresponding flavor-singlet quark helicity PDF is given by \cite{Kovchegov:2015pbl,Kovchegov:2016weo,Kovchegov:2016zex,Kovchegov:2017jxc,Kovchegov:2017lsr}
\begin{align}
\sum_f [ \Delta q^f (x, Q^2 ) + \Delta {\bar q}^f (x, Q^2 ) ] = \sum_f  \int d^2 k_T \, g_{1L}^S (x, k_T^2) = \frac{N_c \, N_f}{2 \pi^3} \, \int\limits_{\Lambda^2/s}^1 \frac{d z}{z}  \, \int\limits_\frac{1}{z \, s}^\frac{1}{z \, Q^2} \, \frac{d x_{10}^2}{x_{10}^2} \,  G (x_{10}^2, z),
\end{align}
where
\begin{align}
G (x_{10}^2, z) = \int d^2 \left( \frac{{\un x}_1 + {\un x}_0}{2}\right) \, G_{10} (z)
\end{align}
with $G_{10} = G_{{\un x}_1 , {\un x}_0}$ and ${\un x}_{10} = {\un x}_1 - {\un x}_0$.

For the flavor non-singlet distribution we have to {\sl subtract} the antiquark contribution out of \eq{TMD19}:
\begin{align}\label{TMD23}
& g_{1L}^{NS} (x, k_T^2) = \frac{4 p^+ i}{(2\pi)^4} \:  \int d^{2} \zeta 
\, d^2 w \, e^{ - i {\un k} \cdot (\ul{\zeta} - \ul{w})} \, \int\limits_0^\infty \frac{d k_1^-}{2\pi} \,  \frac{\un{k} }{\un{k}^2} \cdot  \frac{\un{\zeta} - \un{w}}{|\un{\zeta} - \un{w}|^2}  \\ & \times \,  \left\langle \mbox{T} \, \mbox{tr} \left[ V_{\ul \zeta} \, V_{{\un w}}^{pol \, \dagger} \right] - \mbox{T} \, \mbox{tr} \left[ V_{{\un w}}^{pol} \, V_{\ul \zeta}^\dagger \right] + \bar{\mbox{T}} \, \mbox{tr} \left[ V_{\un \zeta}^{pol} \, V_{\ul w}^\dagger \right]  - \bar{\mbox{T}} \, \mbox{tr} \left[ V_{\ul w} \, V_{\un \zeta}^{pol \, \dagger} \right]  \right\rangle . \notag 
\end{align}
Similarly, define the flavor non-singlet polarized dipole amplitude
\begin{align}\label{GdefwzNS}
G_{{\un w}, {\un \zeta}}^{NS} (zs) = \frac{k_1^- \, p^+}{N_c}  \: \mbox{Re} \,  \left\langle \mbox{T} \, \mbox{tr} \left[ V_{\ul \zeta} \, V_{{\un w}}^{pol \, \dagger} \right] - \mbox{T} \, \mbox{tr} \left[ V_{{\un w}}^{pol} \, V_{\ul \zeta}^\dagger \right] \right\rangle 
= \frac{k_1^- \, p^+}{N_c}  \:  \mbox{Re} \, \left\langle \mbox{tr} \left[ V_{\ul \zeta} \, V_{{\un w}}^{pol \, \dagger} \right] - \mbox{tr} \left[ V_{\ul \zeta}^\dagger \, V_{{\un w}}^{pol} \right] \right\rangle ,
\end{align}
where we have used \eq{Wilson_rels_a} and the complex conjugate of \eq{Wilson_rels_b} to simplify the definition. Using \eq{GdefwzNS}
in \eq{TMD23} we arrive at
\begin{align}\label{TMD24}
g_{1L}^{NS} (x, k_T^2) =  \frac{8 N_c}{(2\pi)^6}  \: \int d^{2} \zeta 
\, d^2 w \, d^2 y \, e^{ - i {\un k} \cdot (\ul{\zeta} - \ul{y})} \, \int\limits_{\Lambda^2/s}^1 \frac{d z}{z}  \,\frac{\un{\zeta} - \un{w}}{|\un{\zeta} - \un{w}|^2} \cdot  \frac{\un{y} - \un{w}}{|\un{y} - \un{w}|^2} \, G^{NS}_{{\un w}, {\un \zeta}} (zs) ,
\end{align}
in agreement with Eq.~(54c) in \cite{Kovchegov:2016zex}. Once again, the apparent difference between the definition of the flavor non-singlet distribution in Eq.~(55a) of \cite{Kovchegov:2016zex} and \eq{GdefwzNS} is due to the real-part (Re) operator and the time-ordering signs which were implied in \cite{Kovchegov:2016zex}, though not shown explicitly. Only cut diagrams were calculated in \cite{Kovchegov:2016zex} for the initial condition of the non-singlet polarized dipole evolution. Another reason for this absence of the Re sign causing no difference in that particular case is that the expression under the Re sign was already real (see Appendix~\ref{B} for details).

%%%%%%%%%%%%%%%%%%%%%%%%%%%%%%%%%%%%%%%%%%%%%%%%%%%%%%%%%%%%%%%%%

\subsection{Gluon Helicity TMDs}

For completeness, let us quote here the results of \cite{Kovchegov:2017lsr}, where the dipole and Weizs\"{a}cker-Williams
(WW) gluon helicity TMDs were calculated at small $x$, also starting with the full operator expression.

Define another dipole-like polarized operator
\begin{align}  \label{eq:Gidef}
G^i_{10} (z s) \equiv \frac{1}{2 N_c} \,  
\left\langle \tr \left[ V_{\ul 0} (V_{\ul 1}^{pol \, \dagger})_\bot^i
\right] + \cc \right\rangle (z s)
\end{align}
with a different polarized fundamental Wilson line 
\begin{align} \label{M:Vpol2} (V_{\ul x}^{pol})_\bot^i &\equiv
  \int\limits_{-\infty}^{+\infty} dx^- \, V_{\ul x} [+\infty, x^-] \:
  \left( i g \, P^+ \, A_\bot^i (x) \right) \: V_{\ul x} [x^- ,
  -\infty]
  \notag \\ &= \half \int\limits_{-\infty}^{+\infty} dx^- \, V_{\ul x}
  [+\infty, x^-] \: \left( i g \, \bar{A}_\bot^i (x) \right) \:
  V_{\ul x} [x^- , -\infty] .
\end{align}

Applying \eq{Wilson_rels_a} we can rewrite \eq{eq:Gidef} as
\begin{align}  \label{eq:Gidef2}
G^i_{10} (z s) \equiv \frac{1}{2 N_c} \,  
\left\langle \mbox{T} \, \tr \left[ V_{\ul 0} (V_{\ul 1}^{pol \, \dagger})_\bot^i
\right] + \cc \right\rangle (z s) ,
\end{align}
which facilitates its diagrammatic evaluation performed in \cite{Kovchegov:2017lsr}.

After the integration over all impact parameters, the new polarized
dipole amplitude is a vector-valued function of $\ul{x}_{10}$ only, with no other transverse vector present. We thus write \cite{Kovchegov:2017lsr}
\begin{align}\label{decomp}
  \int d^2 b_{10} \, G^{i}_{10} (z s) = (x_{10})_\bot^i \, G_1
  (x_{10}^2, z s) + \epsilon_T^{ij} \, (x_{10})_\bot^j \, G_2
  (x_{10}^2, z s) .
\end{align} 

Employing these quantities we write the dipole gluon helicity TMD at small $x$ as \cite{Kovchegov:2017lsr}
\begin{align} 
  \label{M:Gdipcoord}
  g_{1L}^{G \, dip} (x, k_T^2) =
  \frac{- N_c}{\alpha_s \, 2\pi^4} \int d^2 x_{10} \, e^{i \ul{k} \cdot
    \ul{x}_{10}} \, \left[ 1 + x_{10}^2 \frac{\partial}{\partial
      x_{10}^2} \right] G_2 (x_{10}^2 , z s = \tfrac{Q^2}{x}).
\end{align}
The WW gluon helicity TMD is \cite{Kovchegov:2017lsr}
\begin{align} 
  \label{M:GWWcoord} 
  g_{1L}^{G \, WW} (x, k_T^2) &= \frac{1}{\as \, \pi \, (2\pi)^3} \,
   \int d^2 x_{10} \, d^2 b_{10} \: e^{i \un{k}
    \cdot \ul{x}_{10}} \: \epsilon_T^{ij} \: \left\langle \tr \left[
      (V_{\ul 1}^{pol})_\bot^i \, V_{\ul 1}^\dagger \:\: V_{\ul 0}
      \left( \frac{\partial}{\partial (x_0)_\bot^j} V_{\ul 0}^\dagger
      \right) \right] + \cc \right\rangle.
\end{align}
Finally, the gluon helicity PDF is given by
\begin{align} 
  \label{e:coll}
  \Delta G (x, Q^2) &= \int d^2 k \, g_{1L}^{G \, WW} (x, k_T^2) =
  \int d^2 k \, g_{1L}^{G \, dip} (x, k_T^2)
\notag \\ &=
\frac{- 2 N_c}{\alpha_s \pi^2} \left[ \left( 1 + x_{10}^2
    \frac{\partial}{\partial x_{10}^2} \right) \: G_2 (x_{10}^2 , z s
  = \tfrac{Q^2}{x}) \right]_{x_{10}^2 = \tfrac{1}{Q^2}} .
\end{align}

%%%%%%%%%%%%%%%%%%%%%%%%%%%%%%%%%%%%%%%%%%%%%%%%%%%%%%%%%%%%%%%%%

\section{Polarized ``Wilson lines": Operator Definitions}
\label{sec:polW}

%%%%%%%%%%%%%%%%%%%%%%%%%%%%%%%%%%%%%%%%%%%%%%%%%%%%%%%%%%%%%%%%%

\subsection{Polarized fundamental ``Wilson line"}
\label{sec:polWfund}

\begin{figure}[ht]
\begin{center}
\includegraphics[width=  \textwidth]{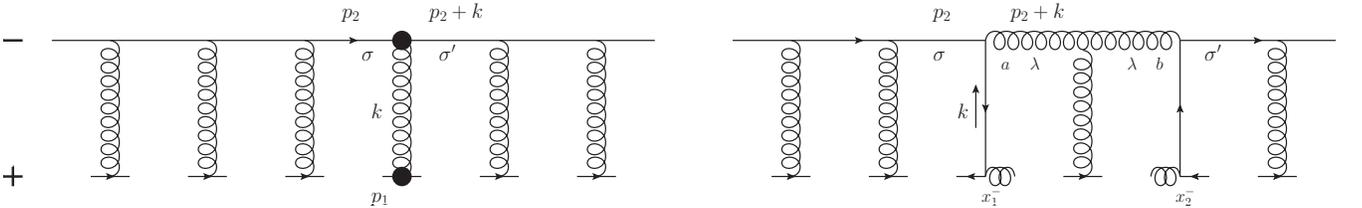} 
\caption{Two contributions to the polarized fundamental Wilson line in a background field. Filled circles at the quark-gluon vertices
  denote the spin-dependent sub-eikonal scattering.}
\label{vpol}
\end{center}
\end{figure}

Our next goal is to construct an explicit expression for the polarized Wilson line operator $V_{\un x}^{pol}$ that we have employed above and in \cite{Kovchegov:2015pbl,Kovchegov:2016weo,Kovchegov:2016zex,Kovchegov:2017jxc,Kovchegov:2017lsr}. To find this operator we have to calculate the scattering amplitude of a high-energy longitudinally polarized quark on a longitudinally polarized target, keeping only the polarization-dependent part of the interaction with the background gluon and quark fields. There are two contributions in this calculation, as shown in the two panels of \fig{vpol}: polarized gluon (left panel) and quarks (right panel) exchanges. The gluon-exchange contribution in the left panel of \fig{vpol} has already been calculated in \cite{Kovchegov:2017lsr}. Hence all is left to do is to find the contribution of the quark exchanges from the right panel.  

%%%%%%%%%%%%%%%%%%%%%%%%%%%%%%%%%%%%%%%%%%%%%%%%%%%%%%%%%%%%%%%%%%%%%
%%%%%%%%%%%%%%%%%%%%%%%%%%%%%%%%%%%%%%%%%%%%%%%%%%%%%%%%%%%%%%%%%%%%%

The expression for the polarized fundamental ``Wilson line" with a single $t$-channel gluon exchange carrying the polarization information, corresponding to the left panel of \fig{vpol}, is given by Eq.~(21) of \cite{Kovchegov:2017lsr}: 
\begin{align} 
  \label{eq:Wpol_fund2} 
  (V^{pol}_{\un x})^g = \frac{i g p_1^+}{s} \,
  \int\limits_{-\infty}^\infty d x^- \, V_{\ul x} [+\infty, x^-] \: 
  F^{12} (x^-, \un{x}) \: V_{\ul x} [x^- , -\infty]. 
\end{align}
Our aim now is to find $(V^{pol}_{\un x})^q$. Let us repeat the calculation from Sec.~II B of \cite{Kovchegov:2017lsr}, but now including quark exchanges in the $t$-channel, as shown in the right panel of \fig{vpol} (exactly what happens to the quark inside the target is not important for our calculation, as long as the target generates the quark fields $\psi$ or $\bar \psi$).  To do so, let us first calculate the contribution of the left $t$-channel quark exchange in the right panel of \fig{vpol},
\begin{align}\label{op1}
\frac{1}{2 p_2^-} \, \epsilon_\lambda^{\mu \, *} (p_2+k) \, (i g) \, {\bar \psi} (k) t^a \gamma_\mu u_\sigma (p_2) = - \frac{i g}{\sqrt{\sqrt{2} \, p_2^-}} \, {\bar \psi} (k) \, t^a \rho (\sigma) \, \delta_{\sigma, \lambda}, 
\end{align}
where we have defined the $+ \leftrightarrow -$ interchanged Brodsky-Lepage spinors $u_\sigma (p_2) = \sqrt{\sqrt{2} \, p_2^-} \, \rho (\sigma)$ for massless quarks with momentum $p_2^\mu = (0, p_2^-, {\un 0}) $ (cf. \cite{Lepage:1980fj}). Here
\begin{align}
  \rho (+1) \, = \, \frac{1}{\sqrt{2}} \, \left(
  \begin{array}{c}
       1 \\ 0 \\ -1 \\ 0
  \end{array}
\right), \ \ \ \rho (-1) \, = \, \frac{1}{\sqrt{2}} \, \left(
  \begin{array}{c}
       0 \\ 1 \\ 0 \\ 1
  \end{array}
\right) ,
\end{align}
and we neglected terms further suppressed by $1/p_2^-$. Fourier transforming \eqref{op1} we get
\begin{align}\label{op2}
- \frac{i g}{\sqrt{\sqrt{2} \, p_2^-}} \, {\bar \psi} (x_1^-, {\un x}) \, t^a \rho (\sigma) \, \delta_{\sigma, \lambda}.
\end{align}
Similarly, the contribution of the right $t$-channel exchange of the right panel in \fig{vpol} gives
\begin{align}\label{op3}
- \frac{i g}{\sqrt{\sqrt{2} p_2^-}} \, \rho^T (\sigma') \, t^b \, \gamma^0 \, {\psi} (x_2^-, {\un x})  \, \delta_{\sigma', \lambda}.
\end{align}
Combining Eqs.~\eqref{op2} and \eqref{op3} we write the operator, the $\sigma \, \delta_{\sigma \sigma'}$-dependent part of which would give us the polarized Wilson line:
\begin{align} 
  \label{eq:Wpol_fundq0} 
  \sigma \, \delta_{\sigma \sigma'} \, (V^{pol}_{\un x})^q \supset - \frac{g^2 \, p_1^+ \, \sqrt{2}}{s}
  \int\limits_{-\infty}^\infty d x_1^- \, \int\limits_{x_1^-}^\infty d x_2^- \, \sum_\lambda \, & V_{\ul x} [+\infty, x_2^-] \:  \rho^T (\sigma') \, t^b  \, \gamma^0 \,  {\psi} (x_2^-, {\un x})  \, \delta_{\sigma', \lambda} \, U_{\ul x}^{ba} [ x_2^-,  x_1^-] \\ & \times \, {\bar \psi} (x_1^-, {\un x}) \, t^a \rho (\sigma) \, \delta_{\sigma, \lambda} \,
  \: V_{\ul x} [x_1^- , -\infty] \notag \\ = - \delta_{\sigma \sigma'} \, \frac{g^2 \, p_1^+ \, \sqrt{2}}{s}
  \int\limits_{-\infty}^\infty d x_1^- \, \int\limits_{x_1^-}^\infty d x_2^- \, & V_{\ul x} [+\infty, x_2^-] \:  \rho^T (\sigma) \, t^b \, \gamma^0 \,   {\psi} (x_2^-, {\un x})  \, U_{\ul x}^{ba} [ x_2^-,  x_1^-] \notag \\ & \times \, {\bar \psi} (x_1^-, {\un x}) \, t^a \rho (\sigma) \: V_{\ul x} [x_1^- , -\infty] \notag \\ = - \delta_{\sigma \sigma'} \, \frac{g^2 \, p_1^+}{s}
  \int\limits_{-\infty}^\infty d x_1^- \, \int\limits_{x_1^-}^\infty d x_2^- \, & V_{\ul x} [+\infty, x_2^-] \:  t^b \, {\psi}_\beta (x_2^-, {\un x})  \, U_{\ul x}^{ba} [ x_2^-,  x_1^-] \left[ \frac{1}{2} \, \gamma^+ \, (1 + \sigma \, \gamma^5) \right]_{\alpha\beta}\notag \\ & \times \, {\bar \psi}_\alpha (x_1^-, {\un x}) \, t^a  \: V_{\ul x} [x_1^- , -\infty] . \notag 
\end{align}

Keeping only the $\sigma$-dependent part of the obtained expression we write
\begin{align} \label{eq:Wpol_fundq} 
(V^{pol}_{\un x})^q = - \frac{g^2 \, p_1^+}{s}
  \int\limits_{-\infty}^\infty d x_1^- \, \int\limits_{x_1^-}^\infty d x_2^- \, & V_{\ul x} [+\infty, x_2^-] \:  t^b \, {\psi}_\beta (x_2^-, {\un x})  \, U_{\ul x}^{ba} [ x_2^-,  x_1^-] \left[ \frac{1}{2} \, \gamma^+ \, \gamma^5 \right]_{\alpha\beta} \, {\bar \psi}_\alpha (x_1^-, {\un x}) \, t^a  \: V_{\ul x} [x_1^- , -\infty] .
\end{align}

Combining Eqs.~\eqref{eq:Wpol_fundq} and \eqref{eq:Wpol_fund2} we finally write the full polarized fundamental ``Wilson line" operator as
\begin{align} 
  \label{eq:Wpol_all} 
  & V^{pol}_{\un x} = \frac{i g p_1^+}{s} \,
  \int\limits_{-\infty}^\infty d x^- \, V_{\ul x} [+\infty, x^-] \: 
  F^{12} (x^-, \un{x}) \: V_{\ul x} [x^- , -\infty] \\ & - \frac{g^2 \, p_1^+}{s}
  \int\limits_{-\infty}^\infty d x_1^- \, \int\limits_{x_1^-}^\infty d x_2^- \, V_{\ul x} [+\infty, x_2^-] \:  t^b \, {\psi}_\beta (x_2^-, {\un x})  \, U_{\ul x}^{ba} [ x_2^-,  x_1^-] \left[ \frac{1}{2} \, \gamma^+ \, \gamma^5 \right]_{\alpha\beta} \, {\bar \psi}_\alpha (x_1^-, {\un x}) \, t^a  \: V_{\ul x} [x_1^- , -\infty]. \notag 
\end{align}

%%%%%%%%%%%%%%%%%%%%%%%%%%%%%%%%%%%%%%%%%%%%%%%%%%%%%%%%%%%%%%%%%

\subsection{Polarized adjoint ``Wilson line"}
\label{sec:polWadj}

Let us now repeat the above calculation (along with the calculation from Sec.~II B of \cite{Kovchegov:2017lsr}), but for the adjoint (gluon) polarized Wilson line. Similar to the above, we have to find the high-energy longitudinally-polarized gluon scattering amplitude on a longitudinally-polarized target, keeping only the polarization-dependent part of the expression.  

%%%%%%%%%%%%%%%%%%%%%%%%%%%%%%%%%%%%%%%%%%%%%%%%%%%%%%%%%%%%%%%%%%%%%
\begin{figure}[ht]
\begin{center}
\includegraphics[width= 0.95 \textwidth]{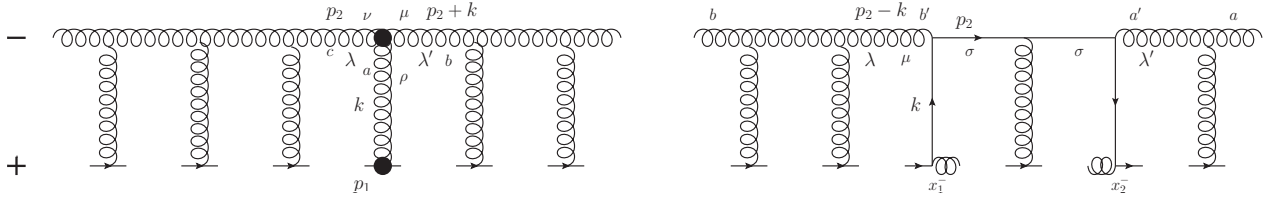} 
\caption{Two contributions to the polarized adjoint Wilson line in the
  quasi-classical approximation (in $A^-=0$ gauge). The filled circles
  denote the spin-dependent sub-eikonal scattering.}
\label{upol}
\end{center}
\end{figure}
%%%%%%%%%%%%%%%%%%%%%%%%%%%%%%%%%%%%%%%%%%%%%%%%%%%%%%%%%%%%%%%%%%%%%

We begin by considering the scattering amplitude in the left panel of \fig{upol}. By analogy to the calculation in \cite{Kovchegov:2017lsr} we write
\begin{align} 
  \label{eq:noneik}
  \lambda \, \delta_{\lambda , \lambda'} \: \hat{\mathcal{O}}_{pol}^{g}
  (k) & \equiv \frac{1}{2 p_2^-} \, \epsilon_{\lambda'}^{\mu \, *} (p_2 +k) \, [(p_2 - k)_\mu \, g_{\nu\rho} - (2 p_2 +k)_\rho \, g_{\mu\nu} + (2 k + p_2)_\nu \, g_{\mu\rho} ] \, \epsilon_{\lambda}^\nu (p_2) \, g \, f^{abc} \,  A^{a \, \rho}_\perp (k) \notag \\ & = \lambda \, \delta_{\lambda , \lambda'} \, \frac{g}{p_2^-} \,  \un{k} \times \un{\cal A} (k)
\end{align}
with all the indices as labeled in the left panel of \fig{upol}. Again we only keep the spin-dependent terms proportional to $\lambda \, \delta_{\lambda , \lambda'}$, while ${\cal A}_\mu$ denotes the color matrix $A^{a}_\mu T^a$ with $T^a$ the
adjoint generators of SU($N_c$). Fourier transforming to
coordinate space gives
\begin{align} \label{eq:noneik2} \hat{\mathcal{O}}_{pol}^{g} (x^- ,
  \ul{x}) &\equiv \int \frac{dk^+}{2\pi} \frac{d^2 k}{(2\pi)^2} \,
  e^{- i k^+ x^-} \, e^{i \ul{k} \cdot \ul{x}} \: \left[ \frac{g}{p_2^-} \un{k} \times \un{\cal A} (k) \right]
\notag \\ & =
\frac{2}{s} \, (- i g p_1^+) \: \epsilon_T^{i j} \:
\frac{\partial}{\partial x_\bot^i} \, {\cal A}_\bot^j (x^-, \un{x}) \equiv
\frac{2}{s} \, (- i g p_1^+) \: \un{\nabla} \times \un{\cal A} (x^-,
\un{x}).
\end{align}

We thus obtain the gluon contribution to the polarized adjoint Wilson line
\begin{align} 
  \label{M:UpolF12}
  (U_{\ul x}^{pol})^g = \frac{2 i \, g \, p_1^+}{s}
  \int\limits_{-\infty}^{+\infty} dx^- \: U_{\ul{x}}[+\infty, x^-] \:
  {\cal F}^{12} (x^+ =0 , x^- , \ul{x}) \: U_{\ul{x}} [x^- , -\infty],
\end{align}
where ${\cal F}^{12}$ is the component of the field-strength tensor in the adjoint representation and 
\begin{align}
  U_{\un{x}} [b^-, a^-] = \mathcal{P} \exp \left[ i g
    \int\limits_{a^-}^{b^-} d x^- \, {\cal A}^+ (x^+=0, x^-, {\un x})
  \right]
\end{align}
is the adjoint Wilson line. 

Finally, defining a rescaled gluon field
\begin{align}
  \label{eq:scale_out}
  \un{\cal A} (x^-, \un{x}) = \frac{S_L}{2 p_1^+} \, \un{\bar {\cal A}} (x^-,
  \un{x})
\end{align}
we obtain
\begin{align} 
  \label{eq:Wpol2} 
  (U^{pol}_{\un x})^g & = \frac{1}{s} \,
  \int\limits_{-\infty}^\infty d x^- \, U_{\ul x} [+\infty, x^-] \:
  \left( - i g \, \epsilon_T^{i j} \: \frac{\partial}{\partial
      x_\bot^i} \, \bar{{\cal A}}_\bot^j (x^-, \un{x}) \right) \: U_{\ul x}
  [x^- , -\infty] \\ & = \frac{i g}{s} \,
  \int\limits_{-\infty}^\infty d x^- \, U_{\ul x} [+\infty, x^-] \:
  \bar{{\cal F}}^{12} (x^-, \un{x}) \: U_{\ul x} [x^- , -\infty]. \notag
\end{align}

Now let us consider the contribution of quark $t$-channel exchanges, as shown in the right panel of \fig{upol}. Starting with the exchange on the left we write for it
\begin{align}
\frac{1}{2 p_2^-} \, i g \, {\bar u}_\sigma (p_2) \, t^{b'} \, \gamma^\mu \, \psi (k) \, \epsilon^\mu_\lambda (p_2-k) = - \frac{i g}{\sqrt{\sqrt{2} p_2^-}} \, \delta_{\sigma, \lambda} \, \rho^T (\sigma) \, t^{b'} \,  \gamma^0 \, \psi (k).
\end{align}

In coordinate space we have
\begin{align}
- \frac{i g}{\sqrt{\sqrt{2} \, p_2^-}} \, \delta_{\sigma, \lambda} \, \rho^T (\sigma) \, t^{b'} \, \gamma^0 \, \psi (x_1^-, {\un x}).
\end{align}
The right $t$-channel quark exchange in the right panel of \fig{upol} yields
\begin{align}
- \frac{i g}{\sqrt{\sqrt{2} \, p_2^-}} \, \delta_{\sigma, \lambda'} \,  {\bar \psi} (x_2^-, {\un x}) \, t^{a'} \,  \rho (\sigma).
\end{align}
Combining these results together we write
\begin{align}
& \lambda \, \delta_{\lambda \lambda'} \, (U^{pol}_{\un x})^{q \, ab} \supset - \frac{g^2}{\sqrt{2} \, p_2^-} \, \delta_{\lambda \lambda'} \, \Bigg[  \int\limits_{-\infty}^\infty d x_1^- \, \int\limits_{x_1^-}^\infty d x_2^- \, U^{aa'}_{\un x} [+\infty, x_2^-] \, {\bar \psi} (x_2^-, {\un x}) \, t^{a'} \,  \rho (\lambda) \, V_{\un x} [x_2^-, x_1^-] \\ &  \times \, \rho^T (\lambda) \, t^{b'} \, \gamma^0 \, \psi (x_1^-, {\un x}) \, U^{b'b}_{\un x} [x_1^-, -\infty] + \int\limits_{-\infty}^\infty d x_1^- \, \int\limits^{x_1^-}_{-\infty} d x_2^- \, U^{ab'}_{\un x} [+\infty, x_1^-] \, {\bar \psi} (x_2^-, {\un x}) \, t^{a'} \,  \rho (\lambda) \, V_{\un x} [x_2^-, x_1^-] \notag \\ &  \times \, \rho^T (\lambda) \, t^{b'} \, \gamma^0 \, \psi (x_1^-, {\un x}) \, U^{a'b}_{\un x} [x_2^-, -\infty]  \Bigg]  = - \frac{g^2}{\sqrt{2} \,  p_2^-} \, \delta_{\lambda \lambda'} \, \int\limits_{-\infty}^\infty d x_1^- \, \int\limits_{x_1^-}^\infty d x_2^- \, U^{aa'}_{\un x} [+\infty, x_2^-] \, \Bigg[ {\bar \psi} (x_2^-, {\un x}) \, t^{a'} \,  \rho (\lambda) \, V_{\un x} [x_2^-, x_1^-] \notag \\ &  \times \, \rho^T (\lambda) \, t^{b'} \, \gamma^0 \, \psi (x_1^-, {\un x}) + c.c. \Bigg]\, U^{b'b}_{\un x} [x_1^-, -\infty] \notag
\end{align}
where the second term in the brackets is due to the contribution of the diagram in which the quark particle number flows in an opposite direction from that in the right panel of \fig{upol}. Simplifying further we arrive at
\begin{align}
\lambda \, \delta_{\lambda \lambda'} \, (U^{pol}_{\un x})^{q \, ab} \supset - \frac{g^2}{2 p_2^-} \, \delta_{\lambda \lambda'} \, \int\limits_{-\infty}^\infty d x_1^- \, \int\limits_{x_1^-}^\infty d x_2^- \, U^{aa'}_{\un x} [+\infty, x_2^-] \, \Bigg\{ & {\bar \psi}_\alpha (x_2^-, {\un x}) \, t^{a'} \, V_{\un x} [x_2^-, x_1^-] \, \left[ \frac{1}{2} \, \gamma^+ (1 + \lambda \, \gamma_5 )\right]_{\alpha \beta} \notag \\ &  \times \, t^{b'} \,  \psi_\beta (x_1^-, {\un x}) + c.c. \Bigg\} \, U^{b'b}_{\un x} [x_1^-, -\infty].
\end{align}
Finally, keeping only the $\lambda$-dependent term we arrive at the expression for the adjoint polarized ``Wilson line" with quark exchanges in the $t$-channel:
\begin{align}\label{Upolq}
(U^{pol}_{\un x})^{q \, ab} = - \frac{g^2 \, p_1^+}{s} \, \int\limits_{-\infty}^\infty d x_1^- \, \int\limits_{x_1^-}^\infty d x_2^- \, U^{aa'}_{\un x} [+\infty, x_2^-] \, \Bigg\{ & {\bar \psi}_\alpha (x_2^-, {\un x}) \, t^{a'} \, V_{\un x} [x_2^-, x_1^-] \, \left[ \frac{1}{2} \, \gamma^+ \gamma_5 \right]_{\alpha \beta} \notag \\ &  \times \, t^{b'} \,  \psi_\beta (x_1^-, {\un x}) + c.c. \Bigg\} \, U^{b'b}_{\un x} [x_1^-, -\infty].
\end{align}

With the help of Eqs.~\eqref{M:UpolF12} and \eqref{Upolq} we derive the full adjoint polarized ``Wilson line" operator:
\begin{align} 
  \label{M:UpolFull}
  & (U_{\ul x}^{pol})^{ab} = \frac{2 i \, g \, p_1^+}{s}
  \int\limits_{-\infty}^{+\infty} dx^- \: \left( U_{\ul{x}}[+\infty, x^-] \:
  {\cal F}^{12} (x^+ =0 , x^- , \ul{x}) \: U_{\ul{x}} [x^- , -\infty] \right)^{ab} \\ &  - \frac{g^2 \, p_1^+}{s} \, \int\limits_{-\infty}^\infty d x_1^- \, \int\limits_{x_1^-}^\infty d x_2^- \, U^{aa'}_{\un x} [+\infty, x_2^-] \,  {\bar \psi} (x_2^-, {\un x}) \, t^{a'} \, V_{\un x} [x_2^-, x_1^-] \, \frac{1}{2} \, \gamma^+ \gamma_5 \, t^{b'} \,  \psi (x_1^-, {\un x}) \, U^{b'b}_{\un x} [x_1^-, -\infty] + c.c.  . \notag
\end{align}

%%%%%%%%%%%%%%%%%%%%%%%%%%%%%%%%%%%%%%%%%

\section{Small-$x$ Helicity Evolution at Large-$N_c$}
\label{sec:largeNc}

We are now ready to use the polarized Wilson line operators derived above to cross-check the small-$x$ helicity evolution equations derived in \cite{Kovchegov:2015pbl,Kovchegov:2016weo,Kovchegov:2016zex,Kovchegov:2017jxc,Kovchegov:2017lsr}. Those equations close only in the large-$N_c$ and the larger-$N_c \& N_f$ limits. We begin with the large-$N_c$ limit, which is dominated by gluons.

\begin{figure}[th]
\begin{center}
\includegraphics[width= 0.7 \textwidth]{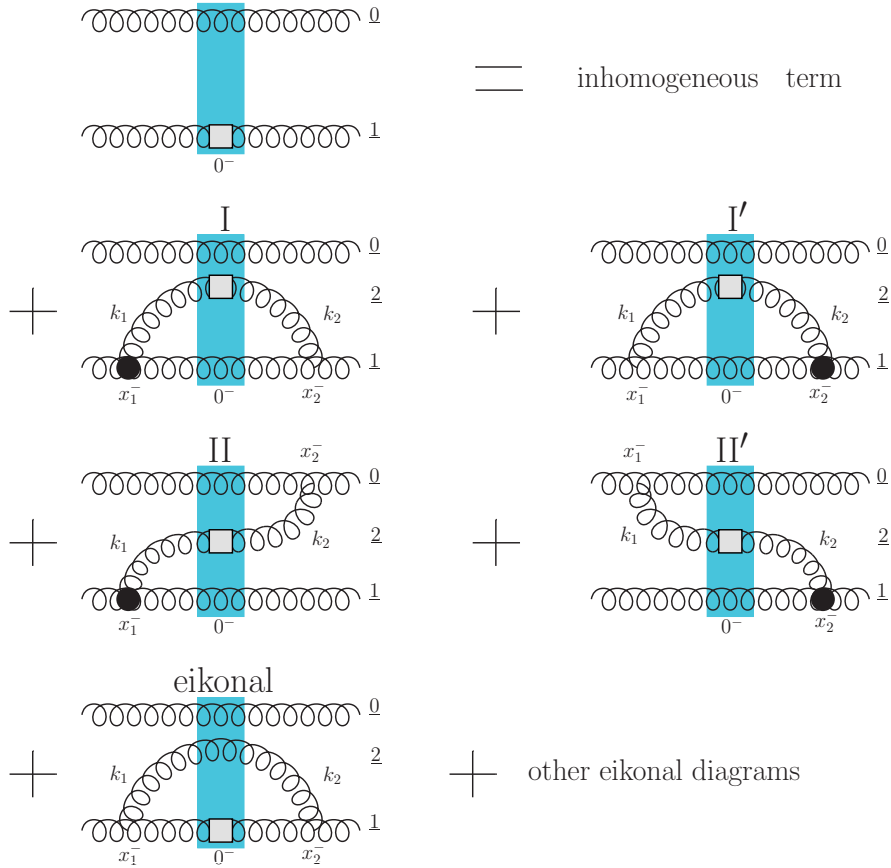} 
\caption{Diagrams illustrating evolution of the polarized dipole amplitude
  \eqref{eq:Gadj_def2}.  The blue rectangle
  represents the target shock wave, the black filled circle
  represents an insertion of the sub-eikonal operator \eqref{eq:noneik2},
  and the gray box represents the polarized adjoint Wilson line \eqref{eq:Wpol2}.}
\label{fig:Gadj_evol_v3}
\end{center}
\end{figure}

We are interested in the evolution of the adjoint polarized dipole amplitude, defined by
\begin{align}\label{eq:Gadj_def2}
G^{adj}_{10} (zs) & = \frac{1}{2 (N_c^2 -1)}  \, \mbox{Re} \, \llangle \mbox{T} \, \mbox{Tr} \left[ U_{\ul 0} \, \left( U_{{\un 1}}^{pol \, g} \right)^\dagger \right] + \mbox{T} \, \mbox{Tr} \left[ U_{{\un 1}}^{pol \, g} \, U_{\ul 0}^\dagger \right] \rrangle \\ & \equiv \frac{z \, s}{2 (N_c^2 -1)}  \, \mbox{Re}  \left\langle \mbox{T} \, \mbox{Tr} \left[ U_{\ul 0} \, \left( U_{{\un 1}}^{pol \, g} \right)^\dagger \right] + \mbox{T} \, \mbox{Tr} \left[ U_{{\un 1}}^{pol \, g} \, U_{\ul 0}^\dagger \right] \right\rangle \notag \\ & = \frac{p^+}{N_c^2 -1} \, 
\int\limits_{-\infty}^\infty d x_1^- \, \mbox{Re} \, \left\langle \mbox{T} \, \mbox{Tr} \left[ U_{\ul
      0} \:\: U_{\ul 1} [-\infty, x_1^-] \, \left( i g \,
      \epsilon_T^{i j} \: \frac{\partial}{\partial (x_1)_\bot^i}
      {\cal A}_\bot^j (x_1^- , \ul{x}_1) \right) \, U_{\ul 1} [x_1^-, \infty]
  \right] + \cc \right\rangle (z s) . \notag  
\end{align}
Here Tr denotes a color trace in the adjoint representation. We are keeping only the gluon operator contribution to the polarized Wilson lines. The diagrams giving the DLA large-$N_c$ evolution of $G^{adj}_{10} (zs)$ are in complete analogy with the Fig.~2 of \cite{Kovchegov:2017lsr}. They are shown in \fig{fig:Gadj_evol_v3} here. In the large-$N_c$ limit quark loops are suppressed. Therefore, the soft quark emission is not included (cf. \fig{fig:Gadj_evol_NcNf} below). For brevity, from now on we will often omit the Re sign, implying that it is applied to all of the correlators below.

%%%%%%%%%%%%%%%%%%%%%%%%%%%%%%%%%%%%%%%%%%%%%%%%%%%%%%%%%%%%%%%%%%%%%
%
%%%%%%%%%%%%%%%%%%%%%%%%%%%%%%%%%%%%%%%%%%%%%%%%%%%%%%%%%%%%%%%%%%%%
%

Employing the notation from \cite{Kovchegov:2017lsr} we write
\begin{align} 
\label{e:qIevol1}
(\delta G^{adj}_{10})_\mathrm{I} = \frac{g^2 \, p^+}{N_c^2 -1}
\, \int\limits_{-\infty}^0 dx_1^-
\int\limits_0^\infty dx_2^- \, \left\langle \mbox{T} \, \Tr\left[ U_{\ul 0} T^a
    U_{\ul 1}^\dagger T^b \right] \:\:
\contraction[2ex]
{\Big( \frac{\partial}{\partial (x_1)_\bot^i} \epsilon_T^{i j} \: }
{a_{\bot}^{j \, a}}
{(x_1^- , \ul{x}_1) \Big) \:}
{a^{+ \, b}}
\Big( \frac{\partial}{\partial (x_1)_\bot^i} \epsilon_T^{i j} \: 
a_{\bot}^{j \, a} (x_1^- , \ul{x}_1) \Big) \:
a^{+ \, b} (x_2^- , \ul{x}_1) 
\:\: + \cc \right \rangle .
\end{align}
With the help of the propagator 
\begin{align} 
\label{e:qIevol2}
\int\limits_{-\infty}^0 dx_1^- 
\int\limits_0^\infty dx_2^- \,
\contraction[2ex]
{\Bigg( \frac{\partial}{\partial (x_1)_\bot^i} \epsilon_T^{i j} \: }
{a_{\bot}^{j \, a}}
{(x_1^- , \ul{x}_1) \Bigg) \:}
{a^{+ \, b}}
\Bigg( \frac{\partial}{\partial (x_1)_\bot^i} &\epsilon_T^{i j} \: 
a_{\bot}^{j \, a} (x_1^- , \ul{x}_1) \Bigg) \:
a^{+ \, b} (x_2^- , \ul{x}_1) 
=
\frac{1}{4\pi^3} \int\limits_0^\infty dk^- \int \frac{d^2
  x_2}{x_{21}^2} \, (U_{\ul 2}^{pol})^{b
  a}_{(k^-)},
\end{align}
which was also established in  \cite{Kovchegov:2017lsr}, we write
\begin{align} 
\label{e:qIevol3} 
(\delta G^{adj}_{10})_{\mathrm{I}} (z s) &= \frac{g^2 p^+}{4 \pi^3 (N_c^2 -1)}
\int\limits_0^\infty dk^- \int \frac{d^2 x_2}{x_{21}^2} \left\langle
  \mbox{T} \, \Tr\left[ U_{\ul 0} T^a U_{\ul 1}^\dagger T^b \right] (U_{\ul
    2}^{pol})^{b a} + \cc \right\rangle (z' s = 2 p^+ k^-)
\notag \\ &=
\frac{\alpha_s}{2 \pi^2} \int\limits_{\frac{\Lambda^2}{s}}^{z}
\frac{dz'}{z'} \int \frac{d^2 x_2}{x_{21}^2} \: \llangle \frac{1}{N_c^2 -1} \,
\mbox{T} \, \Tr\left[ U_{\ul 0} T^a U_{\ul 1}^\dagger T^b \right]
(U_{\ul 2}^{pol})^{b a} + \cc \rrangle (z' s) .
\end{align}
Adding the contribution of the diagram I' simply doubles the result, yielding
\begin{align} 
\label{e:qIevol4} 
(\delta G^{adj}_{10})_{\mathrm{I} + \mathrm{I}'} (z s) &= 
\frac{\alpha_s}{2 \pi^2} \int\limits_{\frac{\Lambda^2}{s}}^{z}
\frac{dz'}{z'} \int \frac{d^2 x_2}{x_{21}^2} \: \llangle \frac{2}{N_c^2 -1} \,
\mbox{T} \, \Tr\left[ U_{\ul 0} T^a U_{\ul 1}^\dagger T^b \right]
(U_{\ul 2}^{pol})^{b a} + \cc \rrangle (z' s) .
\end{align}
The effect of diagrams II and II' in the DLA is to simply introduce the IR cutoff $x_{10} > x_{21}$ on the $x_2$-integral in the diagrams I and I' \cite{Kovchegov:2015pbl,Kovchegov:2016zex}. Finally, the `eikonal' diagrams in \fig{fig:Gadj_evol_v3} are calculated in the same way as for the unpolarized evolution \cite{Balitsky:1995ub,Balitsky:1998ya,Kovchegov:1999yj,Kovchegov:1999ua,Jalilian-Marian:1997dw,Jalilian-Marian:1997gr,Iancu:2001ad,Iancu:2000hn}. Note that the rescaling in the double angle brackets defined in \eq{eq:Gadj_def2} is done with the $z$ of the polarized Wilson line, while the $z$ in the argument of the correlator is the longitudinal momentum fraction of the softest line in the dipole \cite{Kovchegov:2015pbl} (which may be the unpolarized line). In the end we arrive at the following evolution equation for the adjoint polarized dipole amplitude:
\begin{align}\label{e:qIevol30} 
 G^{adj}_{10} (z) = & G^{adj \, (0)}_{10} (z) + \frac{\alpha_s}{2 \pi^2} \int\limits_{\frac{\Lambda^2}{s}}^{z}
\frac{dz'}{z'} \int \frac{d^2 x_2}{x_{21}^2} \: \theta (x_{10} - x_{21}) \:  \theta \left( x_{21}^2 - \frac{1}{z' s} \right) \\ 
& \times \left\{  \llangle \frac{2}{N_c^2 -1} \,
\mbox{T} \, \Tr\left[ U_{\ul 0} T^a U_{\ul 1}^\dagger T^b \right]
(U_{\ul 2}^{pol})^{b a} + \cc \rrangle (z')  \right. \notag \\ & + \left. \frac{1}{N_c^2 -1} \left[
      \left\langle \!\!  \left\langle \mbox{T} \, \mbox{Tr} \left[ T^b \,
            U_{\ul{0}} \, T^a \, U_{\ul{1}}^{pol \, \dagger} \right]
          \, U^{ba}_{\ul{2}} \right\rangle \!\!  \right\rangle (z') -
      N_c \left\langle \!\! \left\langle \mbox{T} \, \mbox{Tr} \left[ U_{\ul{0}}
            \, U_{\ul{1}}^{pol \, \dagger} \right] \right\rangle \!\!
      \right\rangle (z') + \cc  \right] \right\} . \notag 
\end{align}
(We have suppressed the $s$ in $z s$ in the arguments of the functions and correlators in \eqref{e:qIevol30}.)
The evolution equation \eqref{e:qIevol30} is consistent with Eq.~(62) from \cite{Kovchegov:2015pbl} and with Eq.~(A1) in \cite{Kovchegov:2016zex}. 

Next let us take the large-$N_c$ limit of \eq{e:qIevol30}. This means rewriting \eqref{e:qIevol30} in terms of the fundamental polarized dipole amplitudes.  Start with the true (unpolarized) adjoint Wilson line
\begin{align}\label{Uab}
(U_{\ul 0})^{ab} = 2 \, \tr [t^b V_{\ul 0}^\dagger t^a V_{\ul 0}].
\end{align}
To derive a similar relation for $(U^{pol})^g$ from \eq{eq:Wpol2} we write
\begin{align} 
  \label{eq:Wpol3} 
  (U^{pol}_{\un x})^{g \, ab} & = \frac{i g}{s} \,
  \int\limits_{-\infty}^\infty d x^- \, \left( U_{\ul x} [+\infty, x^-] \right)^{ac} \: (T^e)^{cd}
  \: \left( U_{\ul x} [x^- , -\infty] \right)^{db} \,  \bar{F}^{e \, 12} (x^-, \un{x}) \\ & = \frac{4 g}{s} \, \int\limits_{-\infty}^\infty d x^- \, \tr \left[ t^c V^\dagger_{\ul x} [+\infty, x^-] t^a V_{\ul x} [+\infty, x^-] \right] \, f^{ecd} \, \tr \left[ t^b V^\dagger_{\ul x} [x^- , -\infty] t^d V_{\ul x} [x^- , -\infty] \right] \,  \bar{F}^{e \, 12} (x^-, \un{x}) \notag \\ & = \frac{- 8 i g}{s} \, \int\limits_{-\infty}^\infty d x^- \, \tr \left[ t^c V^\dagger_{\ul x} [+\infty, x^-] t^a V_{\ul x} [+\infty, x^-] \right] \, \tr \left[ t^c [t^d, t^e] \right] \, \tr \left[ t^b V^\dagger_{\ul x} [x^- , -\infty] t^d V_{\ul x} [x^- , -\infty] \right] \,  \bar{F}^{e \, 12} (x^-, \un{x})  \notag \\ & = \frac{- 4 i g}{s} \, \int\limits_{-\infty}^\infty d x^- \, \tr \left[ V^\dagger_{\ul x} [+\infty, x^-] t^a V_{\ul x} [+\infty, x^-] \, [t^d, t^e] \right] \, \tr \left[ t^b V^\dagger_{\ul x} [x^- , -\infty] t^d V_{\ul x} [x^- , -\infty] \right] \,  \bar{F}^{e \, 12} (x^-, \un{x})   \notag \\ & = \frac{- 2 i g}{s} \, \int\limits_{-\infty}^\infty d x^- \, \bar{F}^{e \, 12} (x^-, \un{x}) \, \left\{ \tr \left[ t^e V^\dagger_{\ul x} [+\infty, x^-] t^a V_{\ul x} [+\infty, x^-] \, V_{\ul x} [x^- , -\infty] t^b V^\dagger_{\ul x} [x^- , -\infty]  \right] \right. \notag \\ & \left. - \tr \left[ V^\dagger_{\ul x} [+\infty, x^-] t^a V_{\ul x} [+\infty, x^-] t^e \, V_{\ul x} [x^- , -\infty] t^b V^\dagger_{\ul x} [x^- , -\infty]  \right]  \right\} . \notag
\end{align}

The expression for the polarized fundamental ``Wilson line" is given by \eq{eq:Wpol_fund2}, which for the momentum-rescaled gluon field reads 
\begin{align} 
  \label{eq:Wpol_fund} 
  (V^{pol}_{\un x})^g = \frac{i g}{2 \, s} \,
  \int\limits_{-\infty}^\infty d x^- \, V_{\ul x} [+\infty, x^-] \: t^e \, 
  \bar{F}^{e \, 12} (x^-, \un{x}) \: V_{\ul x} [x^- , -\infty]. 
\end{align}
With the help of \eq{eq:Wpol_fund} we rewrite \eq{eq:Wpol3} as (cf. \eq{Uab})
\begin{align} 
  \label{eq:Wpol4} 
  (U^{pol}_{\un x})^{g \, ab} =  4 \, \tr \left[ V^{pol \, \dagger}_{\ul x} t^a V_{\ul x} t^b \right] + 4 \,  \tr \left[ V^\dagger_{\ul x} t^a V_{\ul x}^{pol} t^b \right] .
\end{align}
This is twice larger than Eq.~(A5) in the Appendix~A of \cite{Kovchegov:2016zex}. The latter equation was only conjectured in \cite{Kovchegov:2016zex} and the coefficient in it was not derived or cross-checked. Since (for $N_f=0$) all the evolution equations are linear in $U^{pol}$, our end result (A12) in the same Appendix~A of \cite{Kovchegov:2016zex} would not change from multiplying all $U^{pol}$ in the starting point (A1) in \cite{Kovchegov:2016zex} by a constant. 

Define the fundamental polarized dipole amplitude with only the gluon operator contributing to the fundamental polarized Wilson lines (cf. \eq{Gdefwz})
\begin{align}\label{Ggdef}
 G_{10} (z) = \frac{1}{2 \, N_c} \, \mbox{Re} \, \llangle \mbox{T} \, \tr \left[ V_0 \, (V_1^{pol \, g} )^\dagger \right] + \mbox{T} \,  \tr \left[ V_1^{pol \, g} \, V_0^\dagger \right]   \rrangle .
\end{align}
With the help of \eq{eq:Wpol4} we can see that at large $N_c$
\begin{align}\label{eq:Gadj_Gfund_largeNc}
G^{adj}_{10} (z) = 4 \, G_{10} (z).
\end{align}
The coefficient $4$ of the right of \eq{eq:Gadj_Gfund_largeNc} is twice larger than the more familiar coefficient $2$ in the unpolarized version of this relation.

Repeating all the trace algebra from the Appendix~A of \cite{Kovchegov:2016zex} with Eqs.~\eqref{Uab} and \eqref{eq:Wpol4} defining the unpolarized (normal) and the polarized Wilson lines respectively (instead of (A3) and (A5) of \cite{Kovchegov:2016zex}), yields 
\begin{align}\label{GNc1}
G_{10} (z) = G_{10}^{(0)} (z) + \frac{\alpha_s \, N_c}{2 \pi} \int\limits_{\frac{1}{s \, x_{10}^2}}^{z}
\frac{dz'}{z'} \int\limits^{x_{10}^2}_\frac{1}{z' s} \frac{d x_{21}^2}{x_{21}^2} \: \left[ \Gamma_{10,21} (z') + 3 \, G_{21} (z')  \right],
\end{align}
in complete agreement with the equation derived in \cite{Kovchegov:2015pbl}. (See the next Section for details of this transition.) Here $\Gamma_{10,21}$ is defined operatorially by the same \eq{Ggdef}, but with the dipole size ordering on the subsequent evolution being dependent on the size $x_{21}$ of another dipole \cite{Kovchegov:2015pbl}. We refer to $\Gamma_{10,21}$ as the ``neighbor" dipole amplitude. Its evolution equation is similar to that of $G_{10}$, 
\begin{align}\label{GNc2}
\Gamma_{10,21} (z') = \Gamma_{10,21}^{(0)} (z') + \frac{\alpha_s \, N_c}{2 \pi} \int\limits_{\min \{ \Lambda^2, \frac{1}{x_{10}^2} \} / s }^{z'}
\frac{dz''}{z''} \int\limits^{\min \{ x_{10}^2, x_{21}^2 z'/z'' \} }_\frac{1}{z'' s} \frac{d x_{32}^2}{x_{32}^2} \: \left[ \Gamma_{10,32} (z'') + 3 \, G_{32} (z'')  \right],
\end{align}
and also follows from our operator approach. This result is also in agreement with \cite{Kovchegov:2015pbl}.

In \cite{Kovchegov:2017lsr} we have already constructed the evolution equation for the fundamental polarized dipole amplitude using the operator formalism and anticipating the large-$N_c$ limit. The result is given by Eq.~(74) in \cite{Kovchegov:2017lsr}, which reads
\begin{align} 
  \label{e:qevol1}
  G_{10} (z s) &=
  G_{10}^{(0)} (z s) + \frac{\alpha_s N_c}{2\pi}
  \int\limits^z_{\frac{\Lambda^2}{s}} \frac{dz'}{z'}
  \int\limits^{x_{10}^2}_{1/(z' s)} \frac{d x^2_{21}}{x_{21}^2} \, 
  \Bigg\{ \llangle \frac{1}{N_c^2} \, \mbox{T} \, \tr\left[
    V_{\ul 0} t^a V_{\ul 1}^\dagger t^b \right] (U_{\ul 2}^{pol})^{b
    a} + \cc \rrangle (z' s)
\notag \\ & \hspace{1cm} +
\llangle \frac{1}{N_c^2}
\mbox{T} \,  \tr\left[ V_{\ul 0} t^a V_{\ul 1}^{pol \, \dagger} t^b \right] (U_{\ul
  2})^{b a} - \frac{C_F}{N_c^2} \, \mbox{T} \, \tr\left[ V_{\ul 0} V_{\ul 1}^{pol \,
    \dagger} \right] + \cc \rrangle (z' s) \Bigg\}.
\end{align}

Using \eq{eq:Wpol4} along with  
\begin{align}\label{subst3}
  \left\langle \!\! \left\langle \mbox{tr} \left[ V_{\ul{0}} \, t^a \,
        V_{\ul{1}}^{\dagger} \, t^b \right] \, (U_{\ul 2}^{pol})^{b a}
    \right\rangle \!\! \right\rangle = N_c \, \left\langle
    \!\! \left\langle \mbox{tr} \left[ V_{\ul{0}} \, V^{pol \,
          \dagger}_{\ul{2}} \right] \right\rangle \!\! \right\rangle +
 N_c \, \left\langle \!\! \left\langle \mbox{tr} \left[
        V^{pol}_{\ul{2}} \, V^{\dagger}_{\ul{1}} \right] \right\rangle
    \!\! \right\rangle + {\cal O} \left( \frac{1}{N_c} \right)
\end{align}
we take the large-$N_c$ limit of \eq{e:qevol1} obtaining 
\begin{align} 
  \label{e:qevol300}
  G_{10} (z s) &=
  G_{10}^{(0)} (z s) + \frac{\alpha_s N_c}{2\pi^2}
  \int\limits^z_{\frac{\Lambda^2}{s}} \frac{dz'}{z'}
  \int \frac{d^2 x_2}{x_{21}^2} \, 
  \theta(x_{10}^2 - x_{21}^2) \, \theta (x_{21}^2 - \tfrac{1}{z' s}) \,
  \Bigg\{ \llangle \frac{1}{N_c} \, \mbox{T} \, \mbox{tr}
  \left[ V_{\ul{0}} \, V^{pol \, \dagger}_{\ul{2}} \right] 
\notag \\ & \hspace{1cm} +
   \frac{1}{N_c} \, \mbox{T} \,  \mbox{tr} \left[ V^{pol}_{\ul{2}} \,
    V^{\dagger}_{\ul{1}} \right] + \cc \rrangle (z' s) +
   \llangle \frac{1}{2 N_c} \, \mbox{T} \, \tr\left[ V_{\ul 2} V_{\ul 1}^{pol \,
    \dagger} \right] - \frac{1}{2 N_c} \, \mbox{T} \, \tr\left[ V_{\ul 0} V_{\ul
    1}^{pol \, \dagger} \right] + \cc \rrangle (z' s) \Bigg\},
\end{align}
in agreement with Eq.~(77) from \cite{Kovchegov:2017lsr}. In turn, \eq{e:qevol300} leads to above Eqs.~\eqref{GNc1} and \eqref{GNc2}  for the polarized dipole amplitude and for the neighbor dipole amplitude.

%%%%%%%%%%%%%%%%%%%%%%%%%%%%%%%%%%%%%%%%%%%%%%%%%%%%%%%%%%%%%%%%%

\section{Small-$x$ Helicity Evolution at Large-$N_c \, \& N_f$} 
\label{sec:largeNcNf}

Now let us re-derive helicity evolution equations in the large-$N_c \, \& N_f$ limit using the operators obtained here. Just like in Sec.~\ref{sec:largeNc}, we start with the evolution of the adjoint polarized dipole amplitude, now defined by including the full $U^{pol}$ from \eq{M:UpolFull}:
\begin{align}\label{eq:Gadj_def3}
& G^{adj}_{10} (zs) = \frac{1}{2 (N_c^2 -1)} \, \mbox{Re} \, \llangle \mbox{T} \, \mbox{Tr} \left[ U_{\ul 0} \, U_{{\un 1}}^{pol \, \dagger} \right] + \mbox{T} \, \mbox{Tr} \left[ U_{{\un 1}}^{pol} \, U_{\ul 0}^\dagger \right] \rrangle \notag \\ & = \frac{p^+}{N_c^2 -1} \, 
\int\limits_{-\infty}^\infty d x_1^- \, \mbox{Re} \, \left\langle \mbox{T} \, \mbox{Tr} \left[ U_{\ul
      0} \:\: U_{\ul 1} [-\infty, x_1^-] \, \left( i g \,
      \epsilon_T^{i j} \: \frac{\partial}{\partial (x_1)_\bot^i}
      {\cal A}_\bot^j (x_1^- , \ul{x}_1) \right) \, U_{\ul 1} [x_1^-, \infty]
  \right] + \cc \right\rangle (z s) \notag \\ & - \frac{g^2 \, p^+}{2 (N_c^2 -1)} \, \int\limits_{-\infty}^\infty d x_1^- \int\limits_{x_1^-}^\infty d x_2^- \, \mbox{Re} \,  \left\langle \mbox{T} \, (U_{\ul 0}^\dagger)^{ba} \ U^{aa'}_{\un 1} [+\infty, x_2^-] \, \Bigg\{ {\bar \psi} (x_2^-, {\un x}_1) \, t^{a'} \, V_{\un 1} [x_2^-, x_1^-] \, \frac{1}{2} \, \gamma^+ \, \gamma_5 \, t^{b'} \,  \psi (x_1^-, {\un x}_1) + c.c. \Bigg\} \notag \right. \\ & \times \, \left. U^{b'b}_{\un 1} [x_1^-, -\infty] + \cc \right\rangle (z s) .
\end{align}

One can further simplify the quark contribution to the polarized adjoint Wilson line:
\begin{align}\label{Upolq1}
& (U^{pol}_{\un x})^{q \, ab} = - \frac{g^2 \, p_1^+}{s} \, \int\limits_{-\infty}^\infty d x_1^- \, \int\limits_{x_1^-}^\infty d x_2^- \, U^{aa'}_{\un x} [+\infty, x_2^-] \, \Bigg\{ {\bar \psi} (x_2^-, {\un x}) \, t^{a'} \, V_{\un x} [x_2^-, x_1^-] \, \frac{1}{2} \, \gamma^+ \gamma_5   \, t^{b'} \,  \psi (x_1^-, {\un x}) + c.c. \Bigg\} \, U^{b'b}_{\un x} [x_1^-, -\infty] \notag \\ &  = - \frac{g^2 \, p_1^+}{s} \, \int\limits_{-\infty}^\infty d x_1^- \, \int\limits_{x_1^-}^\infty d x_2^- \, {\bar \psi} (x_2^-, {\un x}) \, V_{\un x} [x_2^-, + \infty] \, t^a \,  V_{\un x} \, \frac{1}{2} \, \gamma^+ \, \gamma_5 \, t^b \, V_{\un x} [-\infty, x_1^-] \, \psi (x_1^-, {\un x}) + c.c.  \\ &  = - \frac{g^2 \, p_1^+}{s} \, \int\limits_{-\infty}^\infty d x_1^- \, \int\limits_{x_1^-}^\infty d x_2^- \,  \tr \left[ t^a \,  V_{\un x} \, t^b \, \left( V_{\un x} [+ \infty, x_2^-] \, \psi (x_2^-, {\un x})_\alpha \, \left( \frac{1}{2} \, \gamma^+ \, \gamma_5 \right)_{\beta\alpha} \, {\bar \psi} (x_1^-, {\un x})_\beta \, V_{\un x} [x_1^-, - \infty]  \right)^\dagger \ \right] + c.c. . \notag
\end{align}

However, it turns out that a relation similar to \eq{eq:Wpol4} connecting the polarized adjoint and fundamental Wilson lines is not easy to obtain in the large-$N_c \, \& N_f$ limit. Instead, we will turn our attention to the polarized dipole amplitudes. Starting with the adjoint amplitude \eqref{eq:Gadj_def3}, we write using the results of the previous Section (after taking the large-$N_c \, \& N_f$ limit in the first term on the right)
\begin{align}\label{eq:Gadj_largeNN1}
& G^{adj}_{10} (z) = 4 \, G_{10} (z)  \notag \\ & - \frac{g^2 \, p^+}{2 (N_c^2 -1)} \, \int\limits_{-\infty}^\infty d x_1^- \, \int\limits_{x_1^-}^\infty d x_2^- \, \Bigg\langle \mbox{T} \, (U_{\ul 0}^\dagger)^{ba} \ U^{aa'}_{\un 1} [+\infty, x_2^-] \, \Bigg\{ {\bar \psi} (x_2^-, {\un x}_1) \, t^{a'} \, V_{\un 1} [x_2^-, x_1^-] \, \frac{1}{2} \, \gamma^+ \, \gamma_5 \, t^{b'} \,  \psi (x_1^-, {\un x}_1) + c.c. \Bigg\} \notag \\ & \times \, U^{b'b}_{\un 1} [x_1^-, -\infty] + \cc \Bigg\rangle (z) ,
\end{align}
where $G_{10} (z)$ is still given by \eq{Ggdef} above. To simplify the second term on the right of \eq{eq:Gadj_largeNN1} we write
\begin{align}
& (U_{\ul 0}^\dagger)^{ba} \ U^{aa'}_{\un 1} [+\infty, x_2^-] \, \Bigg\{ {\bar \psi} (x_2^-, {\un x}_1) \, t^{a'} \, V_{\un 1} [x_2^-, x_1^-] \, \frac{1}{2} \, \gamma^+ \, \gamma_5 \, t^{b'} \,  \psi (x_1^-, {\un x}_1) + c.c. \Bigg\}  \, U^{b'b}_{\un 1} [x_1^-, -\infty] + \cc  \notag \\ & =  U_{\ul 0}^{ab} \ \Bigg\{ {\bar \psi} (x_2^-, {\un x}_1) \, V_{\un 1} [x_2^-, + \infty] \, t^a \,  V_{\un 1} \, \frac{1}{2} \, \gamma^+ \, \gamma_5 \, t^b \, V_{\un 1} [-\infty, x_1^-] \, \psi (x_1^-, {\un x}_1) + c.c. \Bigg\}  + \cc  \notag \\ & = U_{\ul 0}^{ab} \ \Bigg\{ \tr \left[ t^a \,  V_{\un 1} \, t^b \, \left( V_{\un 1} [+ \infty, x_2^-] \, \psi (x_2^-, {\un x}_1)_\alpha \, \left( \frac{1}{2} \, \gamma^+ \, \gamma_5 \right)_{\beta\alpha} \, {\bar \psi} (x_1^-, {\un x}_1)_\beta \, V_{\un 1} [x_1^-, - \infty]  \right)^\dagger \ \right] + c.c. \Bigg\}  + \cc \notag \\ & = \tr \left[ V_{\un 1} \, V_{\un 0}^\dagger \right]  \, \tr \left[ V_{\un 0} \, \left( V_{\un 1} [+ \infty, x_2^-] \, \psi (x_2^-, {\un x}_1)_\alpha \, \left( \frac{1}{2} \, \gamma^+ \, \gamma_5 \right)_{\beta\alpha} \, {\bar \psi} (x_1^-, {\un x}_1)_\beta \, V_{\un 1} [x_1^-, - \infty]  \right)^\dagger \right] + \cc .
\end{align}
Substituting this into \eq{eq:Gadj_largeNN1}, linearizing and taking the large-$N_c \, \& N_f$ limit yields
\begin{align}\label{eq:Gadj_largeNN2}
& G^{adj}_{10} (z) = 4 \, G_{10} (z)  \\ & - \frac{g^2 \, p^+}{2 N_c} \, \int\limits_{-\infty}^\infty d x_1^- \, \int\limits_{x_1^-}^\infty d x_2^- \, \Bigg\langle \mbox{T} \, \tr \left[ V_{\un 0} \, \left( V_{\un 1} [+ \infty, x_2^-] \, \psi (x_2^-, {\un x}_1)_\alpha \, \left( \frac{1}{2} \, \gamma^+ \, \gamma_5 \right)_{\beta\alpha} \, {\bar \psi} (x_1^-, {\un x}_1)_\beta \, V_{\un 1} [x_1^-, - \infty]  \right)^\dagger \right]  + \cc \Bigg\rangle (z) . \notag 
\end{align}

A similar set of operations gives the following expression for the fundamental dipole amplitude (employing the anti-commutativity of the fermion fields):
\begin{align}\label{Qdef}
& Q_{10} (z) \equiv \frac{1}{2 \, N_c} \, \mbox{Re} \, \llangle \mbox{T} \, \tr \left[ V_{\un 0} \, \left( V_{\un 1}^{pol}\right)^\dagger \right] +\mbox{T} \, \tr \left[ V_{\un 1}^{pol} \, V_{\un 0}^\dagger \right]  \rrangle \notag \\ & = G_{10} (z) + \frac{g^2 \, p^+}{4} \, \int\limits_{-\infty}^\infty d x_1^- \, \int\limits_{x_1^-}^\infty d x_2^- \, \left\langle \mbox{T} \, {\bar \psi} (x_1^-, {\un x}_1) \, V_{\un 1} [x_1^-, x_2^-] \, \frac{1}{2} \, \gamma^+ \, \gamma_5 \, \psi (x_2^-, {\un x}_1) + \cc \right\rangle (z). 
\end{align}

Clearly the objects in Eqs.~\eqref{eq:Gadj_largeNN2} and \eqref{Qdef} are significantly different and should obey different evolution equations. 

%%%%%%%%%%%%%%%%%%%%%%%%%%%%%%%%%%%%%%%%%%%%%%%%%%%%%%%%%

\subsection{Adjoint polarized dipole evolution}

%%%%%%%%%%%%%%%%%%%%%%%%%%%%%%%%%%%%%%%%%%%%%%%%%%%%%%%%%%%%%%%%%%%%%
\begin{figure}[ht]
\begin{center}
\includegraphics[width= 0.95 \textwidth]{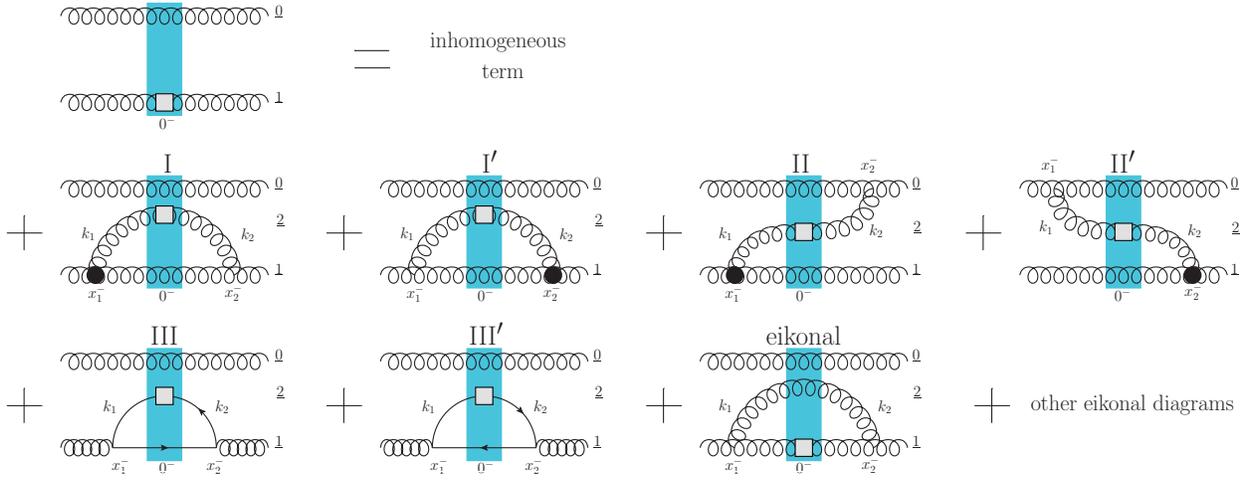} 
\caption{Diagrams illustrating evolution of the polarized dipole amplitude
  \eqref{eq:Gadj_def2} at large-$N_c \, \& N_f$.  The blue rectangle
  represents the classical fields (shock wave), the black vertex
  represents the sub-eikonal operator insertion \eqref{eq:noneik2},
  and the gray box represents the polarized adjoint or fundamental Wilson line.}
\label{fig:Gadj_evol_NcNf}
\end{center}
\end{figure}
%
%%%%%%%%%%%%%%%%%%%%%%%%%%%%%%%%%%%%%%%%%%%%%%%%%%%%%%%%%%%%%%%%%%%%
%

Diagrams contributing to the DLA small-$x$ evolution of the adjoint polarized dipole amplitude in \eq{eq:Gadj_def3} are shown in \fig{fig:Gadj_evol_NcNf}. In comparison with \fig{fig:Gadj_evol_v3} and the large-$N_c$ calculation of the previous Section there are only two new diagrams, the diagrams III and III' in \fig{fig:Gadj_evol_NcNf}. Their contribution is
\begin{align} 
\label{e:qIevol10}
(\delta G^{adj}_{10})_{\mathrm{III} + \mathrm{III'}} = & - \frac{g^2 \, p^+}{2 (N_c^2 -1)}
\, \int\limits_{-\infty}^0 dx_1^-
\int\limits_0^\infty dx_2^-  \\ & \times \, \left\langle \mbox{T} \, (U_{\ul 0})^{ab} \, \Bigg\{ 
\contraction[2ex]
{}{{\bar \psi}}{(x_2^-, {\un x}_1) \, t^{a} \, V_{\un 1} [x_2^-, x_1^-] \, \frac{1}{2} \, \gamma^+ \, \gamma_5 \, t^{b} \,}{\psi}
{\bar \psi} (x_2^-, {\un x}_1) \, t^{a} \, V_{\un 1} [\infty, -\infty] \, \frac{1}{2} \, \gamma^+ \, \gamma_5 \, t^{b} \,  \psi (x_1^-, {\un x}_1) + c.c. \Bigg\} \right\rangle (z s) + \cc . \notag
\end{align}
(The second complex conjugation accounts for the second trace in the polarized dipole amplitude  \eqref{eq:Gadj_def3}: in only doubles the contribution shown explicitly.)
Employing \eq{propagator2} we obtain
\begin{align} 
\label{e:qIevol11}
(\delta G^{adj}_{10})_{\mathrm{III} + \mathrm{III'}} = & - \frac{g^2 \, p^+}{N_c^2 -1}
\, \int\limits_{-\infty}^0 dx_1^-
\int\limits_0^\infty dx_2^-  \, \Bigg\langle \mbox{T} \, (U_{\ul 0})^{ab} \, \left( t^{a} \, V_{\un 1} \, t^{b} \right)^{ij}  \int d^2 w \, \frac{d^2 k_1 \, d k_1^-}{(2\pi)^3 \, (2 k_1^-)^2} \frac{d^2 k_2}{(2\pi)^2} \, e^{i \frac{\un{k}_1^2}{2 k_1^-} x_1^- + i \ul{k}_1 \cdot (\ul{w} - \ul{x}_1)} \\ & \times \, e^{-i  \frac{\un{k}_2^2}{2 k_1^-}  x_2^- + i \ul{k}_2 \cdot (\ul{x}_1 - \ul{w})}  \, \theta (k_1^-) \ \tr \left[ \frac{1}{2} \, \gamma^+ \, \gamma_5 \, \slashed{k_1} \,
\left( \hat{V}_{{\un w}}^\dagger \right)^{ji} \,
\slashed{k_2}  \right]  \Bigg\rangle \Bigg|_{k_2^- = k_1^-, k_1^2 =0, k_2^2 =0} + c.c. . \notag
\end{align}
Integrating over $x_1^-$ and $x_2^-$ yields
\begin{align} 
\label{e:qIevol12}
(\delta G^{adj}_{10})_{\mathrm{III} + \mathrm{III'}} = & \frac{g^2 \, p^+}{N_c^2 -1}
\, \Bigg\langle \mbox{T} \, (U_{\ul 0})^{ab} \, \left( t^{a} \, V_{\un 1} \, t^{b} \right)^{ij}  \int d^2 w \, \frac{d^2 k_1 \, d k_1^-}{(2\pi)^3 \, \un{k}_1^2} \frac{d^2 k_2}{(2\pi)^2 \, \un{k}_2^2} \, e^{i \ul{k}_1 \cdot (\ul{w} - \ul{x}_1)} \, e^{i \ul{k}_2 \cdot (\ul{x}_1 - \ul{w})} \\ & \times \, \theta (k_1^-) \ \tr \left[ \frac{1}{2} \, \gamma^+ \, \gamma_5 \, \slashed{k_1} \, \left( \hat{V}_{{\un w}}^\dagger \right)^{ji} \, \slashed{k_2}  \right]  \Bigg\rangle \Bigg|_{k_2^- = k_1^-, k_1^2 =0, k_2^2 =0} + c.c. . \notag
\end{align}
To evaluate the Dirac matrix trace we have to use polarization sums,
\begin{align}
\tr \left[ \frac{1}{2} \, \gamma^+ \, \gamma_5 \, \slashed{k_1} \, \left( \hat{V}_{{\un w}}^\dagger \right)^{ji} \, \slashed{k_2}  \right] = & \sum_{\sigma_1, \sigma_2} {\bar v}_{\sigma_2} (k_2) \,  \frac{1}{2} \, \gamma^+ \, \gamma_5 \, v_{\sigma_1} (k_1) \, {\bar v}_{\sigma_1} (k_1) \left( \hat{V}_{{\un w}}^\dagger \right)^{ji} \, v_{\sigma_2} (k_2) \\ & =  2 i {\un k}_1 \times {\un k}_2 \, \left( V_{\un w}^\dagger \right)^{ji}  - 2 {\un k}_1 \cdot {\un k}_2 \, \left( V_{\un w}^{pol \, \dagger} \right)^{ji} , \notag
\end{align}
obtaining
\begin{align} 
\label{e:qIevol13}
(\delta G^{adj}_{10})_{\mathrm{III} + \mathrm{III'}} = & - \frac{2 \, g^2 \, p^+}{N_c^2 -1}
\, \Bigg\langle \mbox{T} \, (U_{\ul 0})^{ab} \, \left( t^{a} \, V_{\un 1} \, t^{b} \right)^{ij}  \int d^2 w \, \frac{d^2 k_1 \, d k_1^-}{(2\pi)^3 \, \un{k}_1^2} \frac{d^2 k_2}{(2\pi)^2 \, \un{k}_2^2} \, e^{i \ul{k}_1 \cdot (\ul{w} - \ul{x}_1)} \, e^{i \ul{k}_2 \cdot (\ul{x}_1 - \ul{w})} \\ & \times \, \theta (k_1^-) \ \left[ - i {\un k}_1 \times {\un k}_2 \, \left( V_{\un w}^\dagger \right)^{ji}  + {\un k}_1 \cdot {\un k}_2 \, \left( V_{\un w}^{pol \, \dagger} \right)^{ji} \right]  \Bigg\rangle \Bigg|_{k_2^- = k_1^-, k_1^2 =0, k_2^2 =0} + c.c. . \notag
\end{align}
Fourier-transforming into transverse coordinate space yields
\begin{align} 
\label{e:qIevol14}
(\delta G^{adj}_{10})_{\mathrm{III} + \mathrm{III'}} =  - \frac{\as \, N_f}{2 \, \pi^2} \, \int\limits_{\Lambda^2/s}^1 \frac{dz}{z} \, \int \frac{d^2 w}{|{\un w} - {\un x}_1|^2} \, \frac{1}{N_c^2 -1}
\, \llangle \mbox{T} \, (U_{\ul 0})^{ab} \, \tr \left[ t^{a} \, V_{\un 1} \, t^{b} \, V_{\un w}^{pol \, \dagger} \right] + \bar{\mbox{T}} \, (U_{\ul 0})^{ab} \, \tr \left[ V_{\un w}^{pol} \, t^{b} \, V_{\un 1}^\dagger \, t^a  \right]  \rrangle , 
\end{align}
where we have also inserted a sum over quark flavors. Employing \eq{Uab} we get
\begin{align} 
\label{e:qIevol15}
(\delta G^{adj}_{10})_{\mathrm{III} + \mathrm{III'}} =  - \frac{\as \, N_f}{2 \, \pi^2} \, \int\limits_{\Lambda^2/s}^1 \frac{dz}{z} \, \int \frac{d^2 w}{|{\un w} - {\un x}_1|^2} \, \frac{1}{N_c^2 -1}
\ & \llangle  \half \, \mbox{T} \, \tr \left[ V_{\un 1} \, V_{\un 0}^\dagger \right] \, \tr \left[ V_{\un 0} \, V_{\un w}^{pol \, \dagger}  \right] + \half \, \bar{\mbox{T}} \, \tr \left[ V_{\un 0} \, V_{\un 1}^\dagger \right] \, \tr \left[ V_{\un w}^{pol} \,  V_{\un 0}^\dagger \right] \notag \\ &  - \frac{1}{2 N_c} \, \mbox{T} \, \tr \left[ V_{\un 1} \, V_{\un w}^{pol \, \dagger} \right] - \frac{1}{2 N_c} \, \bar{\mbox{T}} \, \tr \left[ V_{\un w}^{pol} \, V_{\un 1}^\dagger \right]  \rrangle . 
\end{align}
Finally, taking the large-$N_c \, \& N_f$ limit and linearizing the equation by neglecting the LLA evolution (which, in practice, means putting the fundamental traces of unpolarized Wilson lines equal to $N_c$) yields
\begin{align} 
\label{e:qIevol16}
(\delta G^{adj}_{10})_{\mathrm{III} + \mathrm{III'}} =  - \frac{\as \, N_f}{2 \, \pi^2} \, \int\limits_{\Lambda^2/s}^1 \frac{dz}{z} \, \int \frac{d^2 w}{|{\un w} - {\un x}_1|^2} \, {\bar \Gamma}_{{\un w}, {\un 0}; {\un w}, {\un 1}} (z) 
\end{align}
with ${\bar \Gamma}$ being the neighbor polarized dipole amplitude with the polarized line being a true quark, as defined in \cite{Kovchegov:2015pbl}. (Operatorially ${\bar \Gamma}$ is defined by \eq{Qdef}, by analogy with the neighbor dipole amplitude considered above: again, further evolution of ${\bar \Gamma}$ depends on the size of another dipole \cite{Kovchegov:2015pbl}.)

Adding \eq{e:qIevol14} to \eq{e:qIevol30} we obtain
\begin{align}\label{Gevol3all} 
 G^{adj}_{10} (z) = & G^{adj \, (0)}_{10} (z) + \frac{\alpha_s}{2 \pi^2} \int\limits_{\frac{\Lambda^2}{s}}^{z}
\frac{dz'}{z'} \int \frac{d^2 x_2}{x_{21}^2} \: \theta \left( x_{21}^2 - \frac{1}{z' s} \right) \\ 
& \times \left\{ \theta (x_{10} - x_{21}) \:  \llangle \frac{2}{N_c^2 -1} \,
\mbox{T} \, \Tr\left[ U_{\ul 0} T^a U_{\ul 1}^\dagger T^b \right]
(U_{\ul 2}^{pol})^{b a} + \cc \rrangle (z')  \right. \notag \\ & + \theta (x_{10} - x_{21}) \:   \frac{1}{N_c^2 -1} \left[
      \left\langle \!\!  \left\langle \mbox{T} \, \mbox{Tr} \left[ T^b \,
            U_{\ul{0}} \, T^a \, U_{\ul{1}}^{pol \, \dagger} \right]
          \, U^{ba}_{\ul{2}} \right\rangle \!\!  \right\rangle (z') -
      N_c \left\langle \!\! \left\langle \mbox{T} \, \mbox{Tr} \left[ U_{\ul{0}}
            \, U_{\ul{1}}^{pol \, \dagger} \right] \right\rangle \!\!
      \right\rangle (z') + \cc  \right] \notag \\ & \left. - \theta (x_{10}^2 z - x_{21}^2 z') \,
  \frac{N_f}{N_c^2-1} \left\langle \!\! \left\langle \mbox{T} \, \mbox{tr} \left[
        t^b \, V_{\ul{1}} \, t^a \, V_{\ul{2}}^{pol \, \dagger}
      \right] \, U^{ba}_{\ul{0}} + \bar{\mbox{T}} \, \mbox{tr} \left[ t^b \,
        V^{pol}_{\ul{2}} \, t^a \, V_{\ul{1}}^{\dagger} \right] \,
      U^{ba}_{\ul{0}} \right\rangle \!\! \right\rangle (z') \right\}  \notag 
\end{align}
in agreement with Eq.~(A1) of \cite{Kovchegov:2016zex}. 

Using \eq{e:qIevol16} we rewrite this equation as
\begin{align}\label{Gevol3all1}
  & G^{adj}_{10} (z) =  G^{adj \, (0)}_{10} (z) + \frac{\as}{2 \pi^2} \int\limits_{\Lambda^2 /
    s}^z \frac{d z'}{z'} \, \int \frac{d^2 x_{2}}{x_{21}^2} \,
  \theta \left( x_{21}^2 - \frac{1}{z' s} \right) \notag \\ & \times \left\{ \theta (x_{10} - x_{21}) \:  \llangle \frac{2}{N_c^2 -1} \,
\mbox{T} \, \Tr\left[ U_{\ul 0} T^a U_{\ul 1}^\dagger T^b \right]
(U_{\ul 2}^{pol})^{b a} + \cc \rrangle (z')  \right. \notag \\ & \left. + \theta (x_{10} - x_{21}) \:   \frac{1}{N_c^2 -1} \left[
      \left\langle \!\!  \left\langle \mbox{T} \, \mbox{Tr} \left[ T^b \,
            U_{\ul{0}} \, T^a \, U_{\ul{1}}^{pol \, \dagger} \right]
          \, U^{ba}_{\ul{2}} \right\rangle \!\!  \right\rangle (z') -
      N_c \left\langle \!\! \left\langle \mbox{T} \, \mbox{Tr} \left[ U_{\ul{0}}
            \, U_{\ul{1}}^{pol \, \dagger} \right] \right\rangle \!\!
      \right\rangle (z') + \cc  \right] \right\} \notag \\ 
  & - \frac{\as \, N_f}{2 \, \pi} \, \int\limits_{\Lambda^2/s}^z \frac{dz'}{z'} \, \int\limits_{1/(z' \, s)}^{x_{10}^2 z/z'} \frac{d x_{21}^2}{x_{21}^2} \, {\bar \Gamma}_{20; 21} (z'),
\end{align}
where we anticipate the linearized approximation by neglecting LLA terms. Let us reiterate that ${\bar \Gamma}$ is defined as in \eq{Qdef} but for the neighbor dipole amplitude. 

To evaluate the rest of \eq{Gevol3all1} we employ the definition \eqref{M:UpolFull} of the polarized Wilson line. A little algebra involving multiple use of Fierz identity yields
\begin{align}
\frac{1}{N_c^2 -1} \left\langle \!\!
      \left\langle \mbox{T} \, \mbox{Tr} \left[ T^b \, U_{\ul{0}} \, T^a \,
          U_{\ul{1}}^{\dagger} \right] \, \left( U^{pol}_{\ul{2}}
        \right)^{ba} \right\rangle \!\!  \right\rangle (z) = \frac{N_c}{2} \, G_{21}^{adj} (z) + \frac{N_c}{2} \, \Gamma_{20,21}^{adj} (z) + {\cal O} \left( \frac{1}{N_c} \right),
\end{align}
where $\Gamma_{02,21}^{adj}$ is defined just like $G^{adj}$ in \eq{eq:Gadj_def3}, but for the neighbor adjoint dipole. 

Similarly, one can show that
\begin{align}
 \frac{1}{N_c^2 -1} \left\langle \!\!  \left\langle \mbox{T} \, \mbox{Tr} \left[ T^b \,
            U_{\ul{0}} \, T^a \, U_{\ul{1}}^{pol \, \dagger} \right]
          \, U^{ba}_{\ul{2}} \right\rangle \!\!  \right\rangle (z) =  \frac{N_c}{2} \, G_{21}^{adj} (z) + \frac{N_c}{2} \, \Gamma_{10,21}^{adj} (z) + {\cal O} \left( \frac{1}{N_c} \right).
\end{align}

Employing these results in \eq{Gevol3all1} we obtain in the DLA,
\begin{align}\label{Gevol3all3}
   G^{adj}_{10} (z) =  G^{adj \, (0)}_{10} (z) \ + \ & \frac{\as \, N_c}{2 \pi} \int\limits_{\max \{ \Lambda^2 , 1/
    x_{10}^2 \} /s}^z \frac{d z'}{z'} \, \int\limits_{1/(z' \, s)}^{x_{10}^2} \frac{d x_{21}^2}{x_{21}^2} \, \left[ \Gamma_{10,21}^{adj} (z')  + 3 \, G_{21}^{adj} (z')  \right]  \notag \\ 
  & - \frac{\as \, N_f}{2 \, \pi} \, \int\limits_{\Lambda^2/s}^z \frac{dz'}{z'} \, \int\limits_{1/(z' \, s)}^{x_{10}^2 z/z'} \frac{d x_{21}^2}{x_{21}^2} \, {\bar \Gamma}_{10; 21}^{gen} (z'). 
\end{align}
In the last term we have replaced ${\bar \Gamma}_{02; 21} (z')$ by the ``generalized" dipole amplitude (cf. \cite{Kovchegov:2017lsr})
\begin{align}
{\bar \Gamma}_{10; 21}^{gen} (z') = \theta (x_{10} - x_{21}) \, {\bar \Gamma}_{10; 21}^{gen} (z') +  \theta (x_{21} - x_{10}) \, Q_{21} (z'). 
\end{align}
The reason for that is that the neighbor dipole amplitude ${\bar \Gamma}_{20; 21}^{gen} (z')$ is defined (and makes sense in the DLA) only for $x_{21} \gg x_{10} \sim x_{20}$, while the integral in the last term of \eq{Gevol3all3} includes the region where $x_{21} \sim x_{20} \gg x_{10}$: in that region the two neighbor dipoles $21$ and $20$ have comparable sizes and the special neighbor amplitude $\bar \Gamma$ is no longer needed. It has to be replaced by the ``regular" dipole amplitude $Q$, which is accomplished by defining ${\bar \Gamma}^{gen}$. 

Repeating the above steps for the evolution of an adjoint neighbor dipole we arrive at
\begin{align}\label{Gamma_evol_all}
   \Gamma_{10,21}^{adj} (z')  =  \Gamma_{10,21}^{adj  \, (0)} (z') \ + \ & \frac{\as \, N_c}{2 \pi} \int\limits_{\max \{ \Lambda^2 , 1/
    x_{10}^2 \} /s}^{z'} \frac{d z''}{z''} \, \int\limits_{1/(z'' \, s)}^{\min \{ x_{10}^2, x_{21}^2 z'/z'' \} } \frac{d x_{32}^2}{x_{32}^2} \, \left[ \Gamma_{10,32}^{adj} (z'')  + 3 \, G_{32}^{adj} (z'')  \right]  \notag \\ 
  & - \frac{\as \, N_f}{2 \, \pi} \, \int\limits_{\Lambda^2/s}^{z'} \frac{dz''}{z''} \, \int\limits_{1/(z'' \, s)}^{x_{21}^2 z'/z''} \frac{d x_{32}^2}{x_{32}^2} \, {\bar \Gamma}_{10; 32}^{gen} (z''). 
\end{align}

%%%%%%%%%%%%%%%%%%%%%%%%%%%%%%%%%%%%%%%%%%%%%%%%%%%%%%%%%

\subsection{Fundamental polarized dipole evolution}

\begin{figure}[ht]
\begin{center}
\includegraphics[width= 0.95 \textwidth]{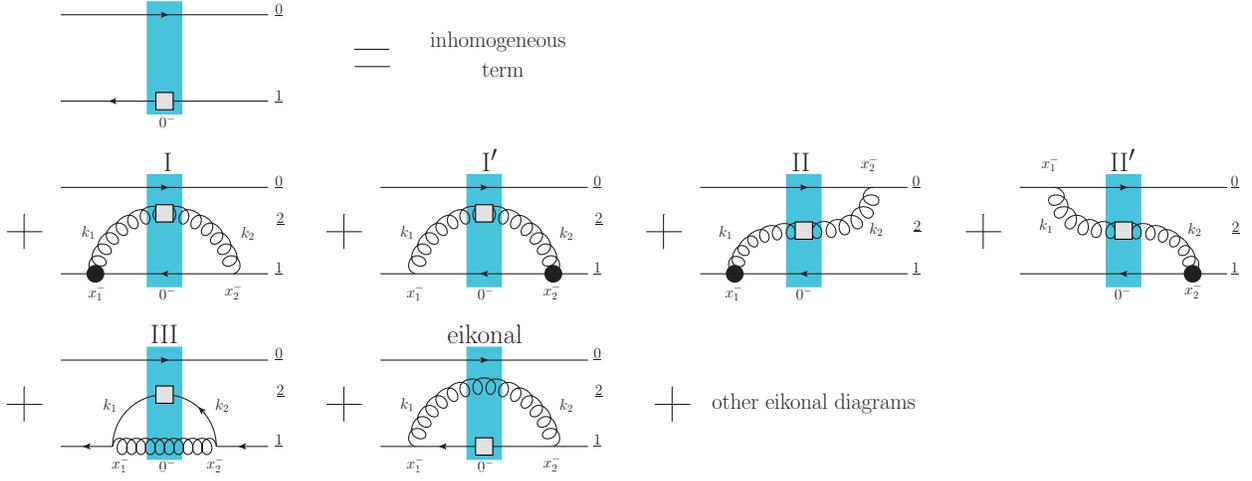} 
\caption{Diagrams illustrating evolution of the polarized dipole amplitude
  \eqref{Qdef} at large-$N_c \, \& N_f$.  The notation is the same as in \fig{fig:Gadj_evol_NcNf}.}
\label{fig:Q_evol_NcNf}
\end{center}
\end{figure}

Next we have to construct the evolution equations for the fundamental polarized dipole amplitude. In the large-$N_c$ limit the evolution equation was constructed in \cite{Kovchegov:2017lsr} and is given above in \eq{e:qevol1}. For the dipole amplitude defined by \eq{Qdef} it reads
\begin{align} 
  \label{e:qevol2}
  Q_{10} (z s) &=
  Q_{10}^{(0)} (z s) + \frac{\alpha_s N_c}{2\pi^2}
  \int\limits^z_{\frac{\Lambda^2}{s}} \frac{dz'}{z'}
  \int \frac{d^2 x_2}{x_{21}^2} \, 
  \theta(x_{10}^2 - x_{21}^2) \, \theta (x_{21}^2 - \tfrac{1}{z' s}) \,
  \Bigg\{ \llangle \frac{1}{N_c^2} \, \mbox{T} \, \tr\left[
    V_{\ul 0} t^a V_{\ul 1}^\dagger t^b \right] (U_{\ul 2}^{pol})^{b
    a} + \cc \rrangle (z' s)
\notag \\ & \hspace{1cm} +
\llangle \frac{1}{N_c^2}
\, \mbox{T} \,  \tr\left[ V_{\ul 0} t^a V_{\ul 1}^{pol \, \dagger} t^b \right] (U_{\ul
  2})^{b a} - \frac{C_F}{N_c^2} \, \mbox{T} \,  \tr\left[ V_{\ul 0} V_{\ul 1}^{pol \,
    \dagger} \right] + \cc \rrangle (z' s) \Bigg\}
\end{align}
corresponding to all the diagrams on the right of \fig{fig:Q_evol_NcNf}, with the exception of diagram III. For large-$N_c \, \& N_f$ limit this equation needs to be augmented by the contribution of the diagram III in \fig{fig:Q_evol_NcNf}. 

%%%%%%%%%%%%%%%%%%%%%%%%%%%%%%%%%%%%%%%%%%%%%%%%%%%%%%%%%%%%%%%%%%%%%
%
%%%%%%%%%%%%%%%%%%%%%%%%%%%%%%%%%%%%%%%%%%%%%%%%%%%%%%%%%%%%%%%%%%%%
%

Using \eq{Qdef} we see that the diagram III in \fig{fig:Q_evol_NcNf} gives
\begin{align}
\left( \delta Q_{10} (z) \right)_\mathrm{III} = \frac{g^2 \, p^+}{4} \, \int\limits_{-\infty}^0 d x_1^- \, \int\limits_{0}^\infty d x_2^- \, \left\langle \, \mbox{T} \, 
\contraction[2ex]
{}{{\bar \psi}}{ (x_1^-, {\un x}_1) \, V_{\un 1} [\infty, -\infty] \, \frac{1}{2} \, \gamma^+ \, \gamma_5 \, }{\psi}
{\bar \psi} (x_1^-, {\un x}_1) \, V_{\un 1} [\infty, -\infty] \, \frac{1}{2} \, \gamma^+ \, \gamma_5 \, \psi 
(x_2^-, {\un x}_1) + \cc \right\rangle (z) .
\end{align}
Evaluating the contraction analogous to the above, we get
\begin{align}
\left( \delta Q_{10} (z) \right)_\mathrm{III} =  \frac{\as}{8 \, \pi^2} \, \int\limits_{\Lambda^2/s}^1 \frac{dz}{z} \, \int \frac{d^2 x_2}{{x}_{21}^2} 
\, \llangle  \mbox{T} \, \tr \left[ V_{\un 1} \, V_{\un 2}^{pol \, \dagger} \right] + \bar{\mbox{T}} \, \tr \left[ V_{\un 2}^{pol} \, V_{\un 1}^\dagger \right]  \rrangle (z) = \frac{\as \, N_c}{4 \, \pi} \, \int\limits_{\Lambda^2/s}^1 \frac{dz}{z} \, \int \frac{d x_{21}^2}{{x}_{21}^2} \, Q_{21} (z).
\end{align}
\eq{e:qevol2}, generalized to the case of the large-$N_c \, \& N_f$ limit, now becomes
\begin{align} 
  \label{e:qevol3}
  Q_{10} (z) & =
  Q_{10}^{(0)} (z s) + \frac{\alpha_s N_c}{2\pi}
  \int\limits^z_{\frac{\Lambda^2}{s}} \frac{dz'}{z'}
  \int\limits_{1/(z' s)}^{x_{10}^2} \frac{d x_{21}^2}{x_{21}^2} \, 
  \Bigg\{ \llangle \frac{1}{N_c^2} \, \mbox{T} \, \tr\left[
    V_{\ul 0} t^a V_{\ul 1}^\dagger t^b \right] (U_{\ul 2}^{pol})^{b
    a} + \cc \rrangle (z')
\\ & +
\llangle \frac{1}{N_c^2}
\, \mbox{T} \, \tr\left[ V_{\ul 0} t^a V_{\ul 1}^{pol \, \dagger} t^b \right] (U_{\ul
  2})^{b a} - \frac{C_F}{N_c^2} \, \mbox{T} \, \tr\left[ V_{\ul 0} V_{\ul 1}^{pol \,
    \dagger} \right] + \cc \rrangle (z') \Bigg\} + \frac{\as \, N_c}{4 \, \pi} \, \int\limits_{\Lambda^2/s}^z \frac{dz'}{z'} \, \int\limits_{1/(z' s)}^{x_{10}^2 z/z'} \frac{d x_{21}^2}{{x}_{21}^2} \, Q_{21} (z') . \notag 
\end{align}

Evaluating the terms on the right-hand side of \eqref{e:qevol3} in the large-$N_c$ and linearized limit we get
\begin{align}\label{relat100}
\llangle \frac{1}{N_c^2} \, \mbox{T} \, \tr\left[
    V_{\ul 0} t^a V_{\ul 1}^\dagger t^b \right] (U_{\ul 2}^{pol})^{b
    a} + \cc \rrangle (z') = \half \, \Gamma_{20,21}^{adj} (z') + \half \, G_{21}^{adj} (z') 
\end{align}
and
\begin{align}\label{relat101}
\llangle \frac{1}{N_c^2} \, \mbox{T} \,
\tr\left[ V_{\ul 0} t^a V_{\ul 1}^{pol \, \dagger} t^b \right] (U_{\ul
  2})^{b a} - \frac{C_F}{N_c^2} \, \mbox{T} \, \tr\left[ V_{\ul 0} V_{\ul 1}^{pol \,
    \dagger} \right] + \cc \rrangle (z') = Q_{21} (z') - {\bar \Gamma}_{10,21} (z'). 
\end{align}
Substituting Eqs.~\eqref{relat100} and \eqref{relat101} back into \eq{e:qevol3} and applying the standard DLA simplifications yields
\begin{align} 
  \label{e:qevol4}
  Q_{10} (z) =
  Q_{10}^{(0)} (z) + & \ \frac{\alpha_s N_c}{2\pi}
  \int\limits^z_{\frac{\Lambda^2}{s}} \frac{dz'}{z'}
  \int\limits_{1/(z' s)}^{x_{10}^2} \frac{d x_{21}^2}{x_{21}^2} \, 
  \Bigg\{ \half \, \Gamma_{10,21}^{adj} (z') + \half \, G_{21}^{adj} (z') 
+ Q_{21} (z') - {\bar \Gamma}_{10,21} (z') \Bigg\} \notag \\ & + \frac{\as \, N_c}{4 \, \pi} \, \int\limits_{\Lambda^2/s}^z \frac{dz'}{z'} \, \int\limits_{1/(z' s)}^{x_{10}^2 z/z'} \frac{d x_{21}^2}{{x}_{21}^2} \, Q_{21} (z') .
\end{align}
Similarly, for the neighbor dipole amplitude we write
\begin{align} 
  \label{e:qevol5}
  {\bar \Gamma}_{10,21} (z') =
  {\bar \Gamma}_{10,21}^{(0)} (z')  + & \ \frac{\alpha_s N_c}{2\pi}
  \int\limits^{z'}_{\frac{\Lambda^2}{s}} \frac{dz''}{z''}
  \int\limits_{1/(z'' s)}^{\min \{ x_{10}^2, x_{21}^2 z'/z'' \} } \frac{d x_{32}^2}{x_{32}^2} \, 
  \Bigg\{ \half \, \Gamma_{10,32}^{adj} (z'') + \half \, G_{32}^{adj} (z'') 
+ Q_{32} (z'') - {\bar \Gamma}_{10,32} (z'') \Bigg\} \notag \\ & + \frac{\as \, N_c}{4 \, \pi} \, \int\limits_{\Lambda^2/s}^{z'} \frac{dz''}{z''} \, \int\limits_{1/(z'' s)}^{x_{21}^2 z/z'} \frac{d x_{32}^2}{{x}_{32}^2} \, Q_{32} (z'') .
\end{align}

%%%%%%%%%%%%%%%%%%%%%%%%%%%%%%%%%%%%%%%%%%%%%%%%%%%%%%%%%

\subsection{Evolution equations at large $N_c \, \& N_f$}

Combining all the above results we write the small-$x$ helicity evolution equations in the large-$N_c \, \& N_f$ limit:
\begin{subequations}\label{all_NcNf}
\begin{align} 
  \label{e:qevol6}
  Q_{10} (z) =
  Q_{10}^{(0)} (z) + & \ \frac{\alpha_s N_c}{2\pi}
  \int\limits^z_{\frac{\Lambda^2}{s}} \frac{dz'}{z'}
  \int\limits_{1/(z' s)}^{x_{10}^2} \frac{d x_{21}^2}{x_{21}^2} \, 
  \Bigg\{ \half \, \Gamma_{10,21}^{adj} (z') + \half \, G_{21}^{adj} (z') 
+ Q_{21} (z') - {\bar \Gamma}_{10,21} (z') \Bigg\} \notag \\ & + \frac{\as \, N_c}{4 \, \pi} \, \int\limits_{\Lambda^2/s}^z \frac{dz'}{z'} \, \int\limits_{1/(z' s)}^{x_{10}^2 z/z'} \frac{d x_{21}^2}{{x}_{21}^2} \, Q_{21} (z') ,
\end{align}
\begin{align}\label{Gevol3all4}
   G^{adj}_{10} (z) =  G^{adj \, (0)}_{10} (z) \ + \ & \frac{\as \, N_c}{2 \pi} \int\limits_{\max \{ \Lambda^2 , 1/
    x_{10}^2 \} /s}^z \frac{d z'}{z'} \, \int\limits_{1/(z' \, s)}^{x_{10}^2} \frac{d x_{21}^2}{x_{21}^2} \, \left[ \Gamma_{10,21}^{adj} (z')  + 3 \, G_{21}^{adj} (z')  \right]  \notag \\ 
  & - \frac{\as \, N_f}{2 \, \pi} \, \int\limits_{\Lambda^2/s}^z \frac{dz'}{z'} \, \int\limits_{1/(z' \, s)}^{x_{10}^2 z/z'} \frac{d x_{21}^2}{x_{21}^2} \, {\bar \Gamma}_{10; 21}^{gen} (z'), 
\end{align}
\begin{align}\label{Gamma_evol_all2}
   \Gamma_{10,21}^{adj} (z')  =  \Gamma_{10,21}^{adj  \, (0)} (z') \ + \ & \frac{\as \, N_c}{2 \pi} \int\limits_{\max \{ \Lambda^2 , 1/
    x_{10}^2 \} /s}^{z'} \frac{d z''}{z''} \, \int\limits_{1/(z'' \, s)}^{\min \{ x_{10}^2, x_{21}^2 z'/z'' \} } \frac{d x_{32}^2}{x_{32}^2} \, \left[ \Gamma_{10,32}^{adj} (z'')  + 3 \, G_{32}^{adj} (z'')  \right]  \notag \\ 
  & - \frac{\as \, N_f}{2 \, \pi} \, \int\limits_{\Lambda^2/s}^{z'} \frac{dz''}{z''} \, \int\limits_{1/(z'' \, s)}^{x_{21}^2 z'/z''} \frac{d x_{32}^2}{x_{32}^2} \, {\bar \Gamma}_{10; 32}^{gen} (z'') , 
\end{align}
\begin{align} 
  \label{e:qevol7}
  {\bar \Gamma}_{10,21} (z') =
  {\bar \Gamma}_{10,21}^{(0)} (z')  \, + & \ \frac{\alpha_s N_c}{2\pi}
  \int\limits^{z'}_{\frac{\Lambda^2}{s}} \frac{dz''}{z''}
  \int\limits_{1/(z'' s)}^{\min \{ x_{10}^2, x_{21}^2 z'/z'' \} } \frac{d x_{32}^2}{x_{32}^2} \, 
  \Bigg\{ \half \, \Gamma_{10,32}^{adj} (z'') + \half \, G_{32}^{adj} (z'') 
+ Q_{32} (z'') - {\bar \Gamma}_{10,32} (z'') \Bigg\} \notag \\ & + \frac{\as \, N_c}{4 \, \pi} \, \int\limits_{\Lambda^2/s}^{z'} \frac{dz''}{z''} \, \int\limits_{1/(z'' s)}^{x_{21}^2 z/z'} \frac{d x_{32}^2}{{x}_{32}^2} \, Q_{32} (z'') .
\end{align}
\end{subequations}

These equations have to be compared to Eqs.~(92) and (93) in \cite{Kovchegov:2015pbl}, while realizing that $A = Q$ there. Writing
\begin{align}
G^{adj} = 2 \, G^{old}, \ \ \ \ \ \Gamma^{adj} = 2 \, \Gamma^{old}
\end{align}
with the $G^{old}, \Gamma^{old}$ denoting the objects used in Eqs.~(92) and (93) of \cite{Kovchegov:2015pbl}, we almost reduce Eqs.~\eqref{all_NcNf} to Eqs.~(92) and (93) of \cite{Kovchegov:2015pbl}. The only remaining difference is due to ${\bar \Gamma}^{gen}$ used in Eqs.~\eqref{Gevol3all4} and \eqref{Gamma_evol_all2} while only $\bar \Gamma$ entered in the similar place in the analogous equations of \cite{Kovchegov:2015pbl}. We believe that our present use of ${\bar \Gamma}^{gen}$ corrects this inaccuracy done in our earlier work.

%%%%%%%%%%%%%%%%%%%%%%%%%%%%%%%%%%%%%%%%%%%%%%%%%%%%%%%%%

\section{Conclusions}
\label{sec:conc}

In this paper we have presented a completely operator-based approach to helicity evolution at small $x$. For the first time ever we have derived explicit expressions for the fundamental and adjoint polarized Wilson lines, given in Eqs.~\eqref{eq:Wpol_all} and \eqref{M:UpolFull} respectively. Employing these expressions, we have re-derived the small-$x$ evolution equations for the polarized dipole amplitudes in the double logarithmic approximation, arriving at Eqs.~\eqref{GNc1} and \eqref{GNc2} in the large-$N_c$ limit and at Eqs.~\eqref{all_NcNf} in the large-$N_c \, \& N_f$ limits. These equations had previously been derived in \cite{Kovchegov:2015pbl} using a combination of the operator-based and diagrammatic methods; Eqs.~\eqref{all_NcNf} contain a minor correction for their prototype in \cite{Kovchegov:2015pbl}.

As mentioned in the Introduction, the large-$N_c$ helicity evolution equations \eqref{GNc1} and \eqref{GNc2} were solved in our earlier works, both numerically \cite{Kovchegov:2016weo} and analytically \cite{Kovchegov:2017jxc}, resulting in the quark helicity asymptotics given in \eq{qG_helicity}. The large-$N_c \, \& N_f$ equations \eqref{all_NcNf} have not been solved yet. Note that, on general grounds, one expects the large-$N_c \, \& N_f$ equations to be more realistic than the large-$N_c$ ones, since the former include the true quark contribution, in addition to the gluon one. Hence we believe solution of the (presently corrected) large-$N_c \, \& N_f$ helicity evolution equations would represent an important next step in our theoretical understanding of quark helicity at small $x$. 

Another important future research direction which may result from the present work is the possibility of obtaining the helicity analogue of JIMWLK evolution, possibly following the method outlined in \cite{Mueller:2001uk} for re-deriving the original unpolarized JIMWLK evolution starting from the evolution of Wilson lines. Obtaining a helicity JIMWLK equation may allow one to numerically determine the small-$x$ asymptotics of quark (and possibly gluon) helicity distributions outside the large-$N_c$ and large-$N_c \, \& N_f$ limits addressed above and in \cite{Kovchegov:2015pbl,Kovchegov:2016weo,Kovchegov:2016zex,Kovchegov:2017jxc,Kovchegov:2017lsr}. In addition, the evolution of higher-order (beyond-dipole amplitude) correlators, such as color-quadrupoles, sextupoles, etc., including exactly one polarized Wilson line may be derived using helicity JIMWLK evolution. 

Finally, while the discussion in this work is dedicated to small-$x$ helicity evolution only, the operator techniques we develop here can be used to determine the small-$x$ asymptotics of other TMDs. The prescription remains the same as above (see also \cite{Kovchegov:2017lsr}): 
\begin{itemize}
\item[(i)] Start with the operator definition of a given TMD and simplify it in the small-$x$ limit. 

\item[(ii)] For quark distribution this results in expressing the TMDs in terms of the polarized Wilson lines, the exact expressions for which have to be determined in a separate calculation. For gluon distribution, the explicit form of the corresponding polarized Wilson lines emerges explicitly from the simplification of the operator definition at small $x$ \cite{Kovchegov:2017lsr}. (The polarized Wilson lines entering the expressions for the quark and gluon distributions at small $x$ were different in the case of helicity, and not only by the color representation factors \cite{Kovchegov:2017lsr}: it is natural to expect that the difference will persist for other spin-dependent quark and gluon TMDs.)

\item[(iii)] Construct the small-$x$ evolution of the polarized dipole operators made out of the obtained fundamental and adjoint polarized Wilson lines. The evolution may start in the DLA limit, if it applies for a given TMD, and continue with the LLA corrections and beyond. Otherwise the evolution may start in the LLA limit, as is the case for unpolarized distributions. (At higher orders impact factors have to be included as well.) The equations are likely to close in the large-$N_c$ and large-$N_c \, \& N_f$ limits only. 

\item[(iv)] Solve the obtained equations, either numerically or, if possible, analytically, to obtain the small-$x$ asymptotics of the TMD in question. 

\end{itemize}

Application of this prescription to the quark transversity distribution is under way and will be reported on shortly \cite{KStransversity}.

%||||||||||||||||||||||||||||||||||||||||||||||||||||||||||||||||||||||||||||||||||||||||||||||||||||||||||||||||||||||||
%||||||||||||||||||||||||||||||||||||||||||||||||||||||||||||||||||||||||||||||||||||||||||||||||||||||||||||||||||||||||
%||||||||||||||||||||||||||||||||||||||||||||||||||||||||||||||||||||||||||||||||||||||||||||||||||||||||||||||||||||||||

\section*{Acknowledgments}

The authors would like to thank Ian Balitsky and Giovanni Chirilli for insisting (for years, in the case of Balitsky) that our helicity evolution equations can and should be rewritten completely in the operator language, with the explicit definition of all the operators.
This material is based upon work supported by the U.S. Department of
Energy, Office of Science, Office of Nuclear Physics under Award
Number DE-SC0004286 (YK) and DOE Contract No. DE-AC52-06NA25396 (MS).  MS
received additional support from the U.S. Department of Energy, Office
of Science under the DOE Early Career Program.\\

%||||||||||||||||||||||||||||||||||||||||||||||||||||||||||||||||||||||||||||||||||||||||||||||||||||||||||||||||||||||||
%||||||||||||||||||||||||||||||||||||||||||||||||||||||||||||||||||||||||||||||||||||||||||||||||||||||||||||||||||||||||
%||||||||||||||||||||||||||||||||||||||||||||||||||||||||||||||||||||||||||||||||||||||||||||||||||||||||||||||||||||||||

%%%%%%%%%%%%%%%%%%%%%%%%%%%%%%%%%%%%%%%%%%%%%%%%%%%%%%%%%%%%%%%%%%%%

\appendix
 
%%%%%%%%%%%%%%%%%%%%%%%%%%%%%%%%%%%%%%%%%%%%%%%%%%%%%%%%%%%%%%%%%%%%

\section{Diagrams A and E cancellation}

%%%%%%%%%%%%%%%%%%%%%%%%%%%%%%%%%%%%%%%%%%%%%%%%%%%%%%%%%%%%%%%%%%%%

 \label{A}

\begin{figure}[ht]
\begin{center}
\includegraphics[width= 0.85 \textwidth]{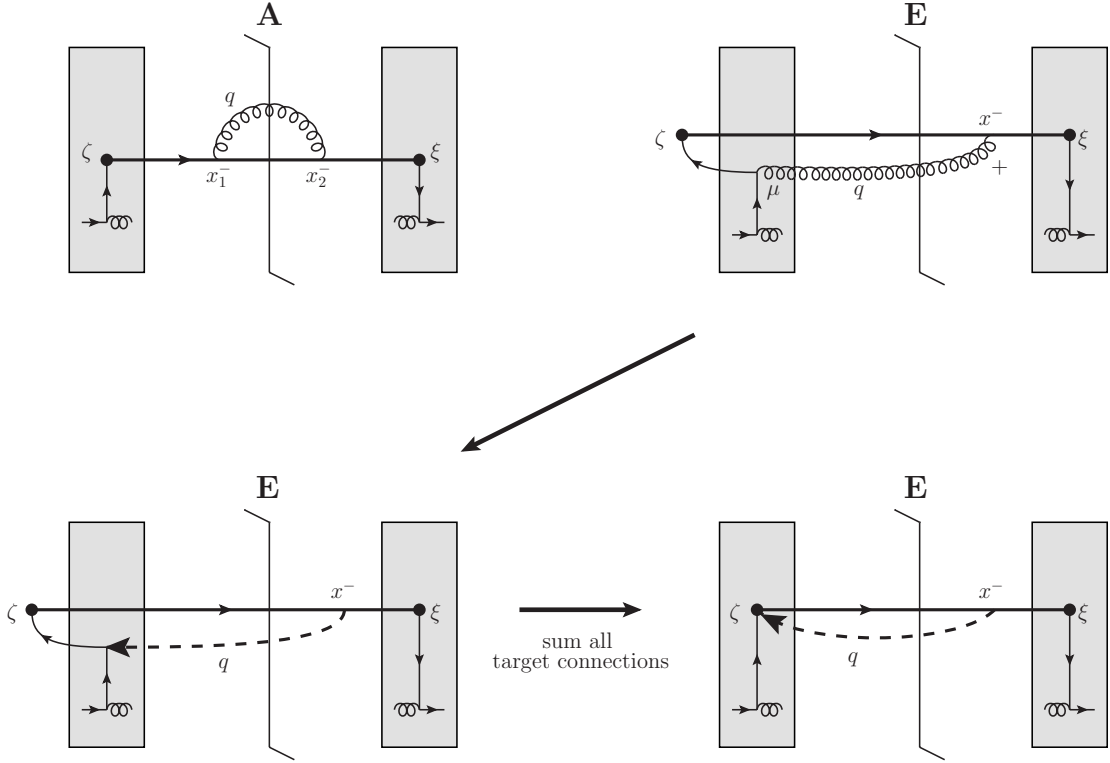} 
\caption{Top row: one-gluon corrections to the diagrams A and E. Bottom row: the leading (DLA) part of the diagram E comes with the longitudinally polarized contribution to the gluon propagator, denoted by the dashed line with the arrow at the end following  \cite{tHooft:1971qjg,Sterman:1994ce}. The diagram in the right panel of the bottom row results from adding to the left bottom-row diagrams all the graphs with all other possible arrow-end of the dashed line connections to the shock wave.}
\label{fig:AE}
\end{center}
\end{figure}

The one-gluon contributions to the diagrams A and E are shown in the top row of \fig{fig:AE}, where instead of the diagram E we are showing it mirror image. 

%%%%%%%%%%%%%%%%%%%%%%%%%%%%%%%%%%%%%%%%%%%%%%%%%%%%%%%%%%%%%%%%%%%%
%
%%%%%%%%%%%%%%%%%%%%%%%%%%%%%%%%%%%%%%%%%%%%%%%%%%%%%%%%%%%%%%%%%%%%

We begin by evaluating the contribution to the diagram A in \fig{fig:AE}. There we only need to evaluate the contribution of the Wilson lines. Working in the $A^- =0$ gauge we write for the Wilson line contribution, mainly arising from the ``++" component of the gluon propagator,
\begin{align}\label{AW1}
(i g ) \, (-i g) \, C_F \int\limits_0^\infty  d x_1^- \, d x_2^- \, e^{- \epsilon (x_1^- + x_2^-)} \, \int \frac{d^4 q}{(2 \pi)^4} \, e^{- i q^+ (x_2^ - - x_1^-) - i {\un q} \cdot ({\un \zeta} - {\un \xi}) } \, (2 \pi) \, \delta^{(+)} (q^2) \, \frac{2 q^+}{q^-} = \frac{2 \, \as \, C_F}{\pi} \, Y \, \ln \frac{1}{|{\un \zeta} - {\un \xi}|  \, \Lambda},
\end{align}
where we have replaced
\begin{align}
\int\limits_{0}^\infty \frac{d q^-}{q^-} \to Y ,
\end{align}
as expected when the $q^-$-integral is properly regularized ($Y$ is the rapidity variable). In \eq{AW1} we have also included exponential regulators for the $x^-$ integrals \cite{Chirilli:2015tea}. With the help of Eqs.~\eqref{AW1} and \eqref{TMD11} we arrive at the following contribution of the diagram A in \fig{fig:AE}:
\begin{align}\label{A1}
A = \frac{2 p^+}{(2\pi)^3} \: \int d^{2} \zeta \, d \zeta^- \, d^{2} \xi \, d \xi^-
\, e^{i k \cdot (\zeta - \xi)} 
\left\langle \bar\psi (\xi) \, \thalf \gamma^+ \gamma^5 \,
 \psi (\zeta) \right\rangle \, \frac{2 \, \as \, C_F}{\pi} \, Y \, \ln \frac{1}{|{\un \zeta} - {\un \xi}| \, \Lambda} .
\end{align}

To evaluate the diagrams in the E-class, first let us note that the contribution of the Wilson line and the gluon propagator is proportional to 
\begin{align}\label{E1}
\int\limits_{0}^\infty d x^- \, e^{- i q^+ x^- - \epsilon x^-} \, (2 \pi) \, \delta^{(+)} (q^2)  \, D^{\mu +} (q) = (2 \pi) \, \delta^{(+)} (q^2)  \, \left[ - g^{\mu +}  + \frac{q^\mu + q^+ \, {\bar \eta}^\mu}{q^-} \right] \, \frac{-i}{q^+ - i \epsilon}
\end{align}
where ${\bar \eta}^\mu = (1, 0, {\un 0})$ in the $(+,-,\perp)$ notation. To obtain a logarithm of energy, we need to have $1/q^-$: hence, the $g^{\mu +}$ term in \eq{E1} can be discarded. Since ${\bar \eta} \cdot v =v^-$ for any 4-vector $v^\mu$, one can show that the term in \eq{E1} containing ${\bar \eta}^\mu$ would lead to a power of the minus momentum of the target proton, which is very small (for the plus-moving proton that we have). Hence this term can also be discarded. We are left with
\begin{align}\label{E1.5}
(2 \pi) \, \delta^{(+)} (q^2) \, \left[  \frac{q^\mu}{q^-} \right] \, \frac{-i}{q^+ - i \epsilon}.
\end{align}
The transition from \eq{E1} to \eq{E1.5} is illustrated in the left panel of the second row of \fig{fig:AE}, where the gluon line is replaced by a dashed line with the arrow indicating the end of the dashed line corresponding to the $q^\mu$ factor in the  part of the propagator left in \eq{E1.5}. This is the standard convention for the gluon lines with a longitudinal polarization on one end that is used to diagrammatically illustrate the Ward identity in QCD \cite{tHooft:1971qjg,Sterman:1994ce}.

%%%%%%%%%%%%%%%%%%%%%%%%%%%%%%%%%%%%%%%%%%%%%%%%%%%%%%%%%%%%%%%%%%%%%
\begin{figure}[ht]
\begin{center}
\includegraphics[width= 0.95 \textwidth]{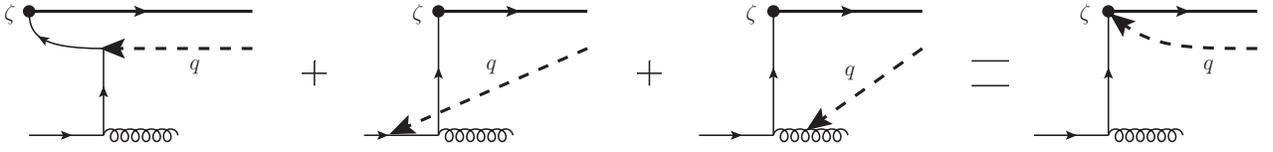} 
\caption{Application of the Ward identity to the diagrams in the E-class. We only depict the amplitude of the diagram: the complex conjugate amplitude remains the same for all graphs.}
\label{fig:ward2}
\end{center}
\end{figure}
%%%%%%%%%%%%%%%%%%%%%%%%%%%%%%%%%%%%%%%%%%%%%%%%%%%%%%%%%%%%%%%%%%%%%

To apply the Ward identity we need to add the diagrams where the dashed line connects (with the arrow end) to the rest of the shock wave in the amplitude. This application of the Ward identity is pictured in \fig{fig:ward2}: summing over all the E-class diagrams where the gluon connects to parts of the target, we arrive at the contribution graphically depicted in the lower-right panel of \fig{fig:AE} in the notation of \cite{tHooft:1971qjg,Sterman:1994ce}. This means that the field $\psi (\zeta)$ remains intact. We thus obtain for the contribution of the E-class diagrams considered here
\begin{align}\label{E2}
E + \ldots = \frac{2 p^+}{(2\pi)^3} \int d^{2} \zeta \, d \zeta^- \, d^{2} \xi \, d \xi^-
\, e^{i k \cdot (\zeta - \xi)} 
\left\langle \bar\psi (\xi) \, \thalf \gamma^+ \gamma^5 \,
 \psi (\zeta) \right\rangle \, (i g ) \, (-ig) \, (-i) \, C_F  \notag \\ \times \, \int \frac{d^4 q}{(2 \pi)^4} \, e^{- i {\un q} \cdot ({\un \zeta} - {\un \xi}) } \, (2 \pi) \, \delta^{(+)} (q^2) \, \frac{1}{q^-} \, \frac{-i}{q^+ - i \epsilon}.
\end{align}
The additional factor of $-i$ arises due to Ward identity. Performing all the $q$-integrals in \eq{E2} we arrive at
\begin{align}\label{E3}
E + \ldots = - \frac{2 p^+}{(2\pi)^3} \int d^{2} \zeta \, d \zeta^- \, d^{2} \xi \, d \xi^-
\, e^{i k \cdot (\zeta - \xi)} 
\left\langle \bar\psi (\xi) \, \thalf \gamma^+ \gamma^5 \,
 \psi (\zeta) \right\rangle \, \frac{\as \, C_F}{\pi} \, Y \, \ln \frac{1}{|{\un \zeta} - {\un \xi}| \, \Lambda}  = - \frac{A}{2}.
\end{align}
We conclude that the E-class diagrams cancel the diagram A. The other half of $A$ is canceled by the complex conjugate of the diagram E in \fig{fig:AE}.\footnote{The cancellation demonstrated here assumes that the range of $q^-$ integrals is the same in the diagrams A and E: while this is correct in the leading logarithmic approximation in $x$, it is not true beyond the small-$x$ approximation, where large logarithms of $Q^2$ are generated in the sum of A and E diagrams \cite{Mueller:2016xoc}, contributing to the Sudakov form factor. Here we assume that $Q^2$ is not large enough to require a separate resummation of $\ln Q^2$.} 

The calculation can be repeated with the same conclusion for the A-type diagram where the gluon line begins and ends on the same Wilson line, say the line that begins at $\zeta$ in  \fig{fig:AE}. In this case we would consider the E-type diagrams with the gluon (dashed) line  \fig{fig:AE} that does not cross the final-state cut, and is emitted by the same Wilson line originating at $\zeta$. The A-type diagram with the gluon line beginning and ending on the same Wilson line has an extra symmetry factor of $1/2$, ensuring the exact cancellation. 

Similar cancellation of diagrams A, E and C is likely to be valid at higher orders in $\as$ and at LLA in $1/x$, by successive application of Ward identity.

%%%%%%%%%%%%%%%%%%%%%%%%%%%%%%%%%%%%%%%%%%%%%%%%%%%%%%%%%%%%%%%%%%%%

\section{Polarized dipole amplitudes at Born level}
\label{B}

Let us compare the calculation of the initial conditions for the flavor singlet and non-singlet polarized dipole amplitudes defined above in Eqs.~\eqref{Gdefwz} and \eqref{GdefwzNS} to what was done in \cite{Kovchegov:2016zex}. For simplicity we will focus on the $t$-channel quark exchanges: the exchanges of $t$-channel gluons can be done similarly. \\

{\sl Flavor-singlet case:} \\

{\bf \oone} First consider the definition in \eq{Gdefwz}. Assuming that the target is a single quark, the flavor-singlet quark helicity TMD is proportional to 
\begin{align}\label{B1}
\mbox{Re} \, \bra{q} \mbox{T} \, \tr \left[ V_0 \, V_1^{pol, \, \dagger} \right] \ket{q} + \mbox{Re} \, \bra{q}  \mbox{T} \, \tr \left[ V_1^{pol}  \, V_0^\dagger \right] \ket{q} = - \thalf \parbox{2.3cm}{\includegraphics[width=2.3cm]{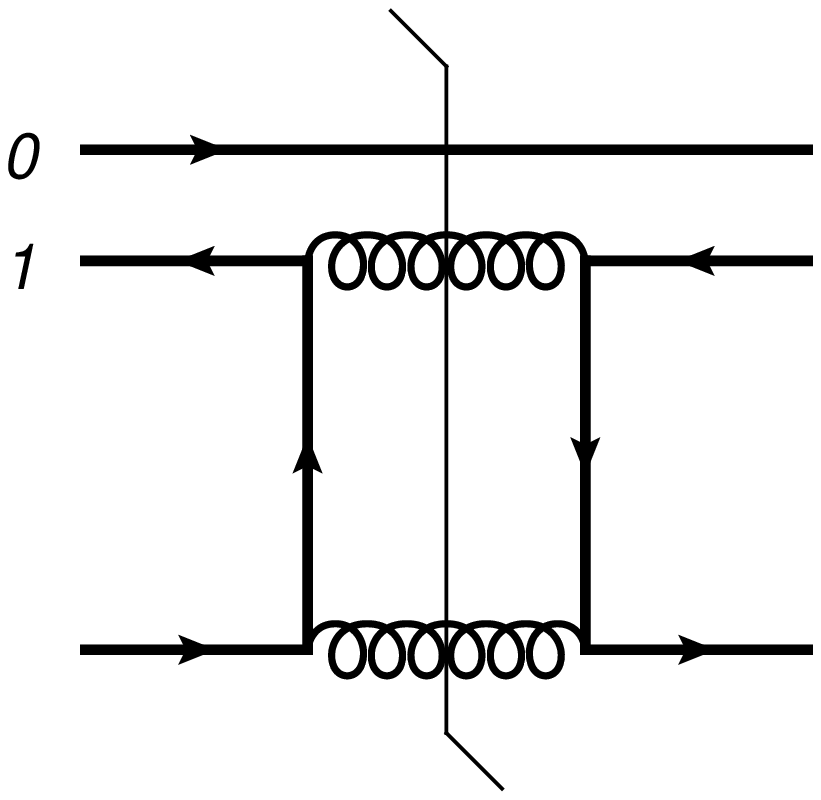}} \ ,
\end{align}
in agreement with Eq.~(10) of \cite{Kovchegov:2016zex}. \\

{\bf \otwo} Now let us try to see what this result means for uncut diagrams. Again assume that the target is a quark. Furthermore, note that the matrix element of the Wilson lines gives us an expectation value of an $S$-matrix (or, for the problem at hand, its spin-dependent part), which is given by $i \, M$ with $M$ the scattering amplitude. That is, 
\begin{align}\label{B3}
i \, M \equiv \bra{q}  \mbox{T} \, \tr \left[ V_0 \, V_1^{pol, \, \dagger} \right] \ket{q} =  \parbox{2.3cm}{\includegraphics[width=2.3cm]{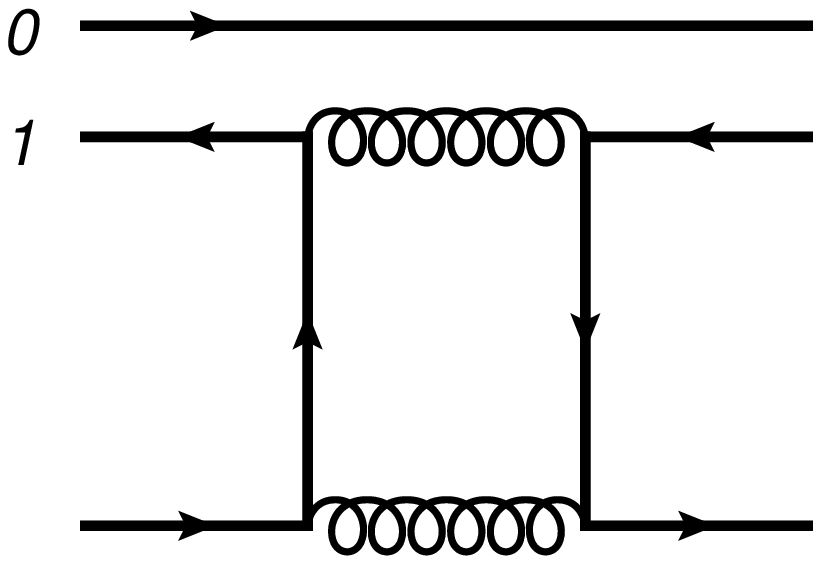}} \ .
\end{align}
(Only the first term in \eq{B1} has a real part for $t$-channel quark exchanges we are restricting ourselves to: hence we will only keep this term in the short exercise below.)

We see that 
\begin{align}\label{B4}
\mbox{Re} [i \, M] = & \, - \mbox{Im} M  = \thalf [ i \, M + (i \, M)^* ] = \thalf \, \bra{q}  \mbox{T} \, \tr \left[ V_0 \, V_1^{pol, \, \dagger} \right] \ket{q} + \thalf \, \bra{q}  \overline{\mbox{T} } \, \tr \left[ V_1^{pol} \, V_0^\dagger \right] \ket{q} \notag \\ & =  \thalf \, \parbox{2.3cm}{\includegraphics[width=2.3cm]{t_quarks2}} \ +  \thalf \, \overline{\mbox{T} } \ \parbox{2.3cm}{\includegraphics[width=2.3cm]{t_quarks2}} \ = -  \thalf \, \parbox{2.3cm}{\includegraphics[width=2.3cm]{t_quarks1}} \ .
\end{align}
Note that under the anti-time ordering $\overline{\mbox{T} }$ the ``regular" Wilson line, say $V_1$, denotes a quark (particle number flows in the same direction as the new time), which for the normal-flowing time would appear as an anti-quark; similarly the conjugate Wilson line $V_0^\dagger$ denotes an anti-quark, but appears to be a quark for the normal-flowing time. The target quark state $\ket{q}$ can be thought of as an anti-quark under $\overline{\mbox{T} }$, since the particle number flows opposite to the new time: if we represent it by another Wilson line operator in the amplitude, it would conjugate in the cc amplitude, giving an anti-quark. Hence, the second diagram in the second line of \eq{B4} looks just like the first diagram, only time in the second diagram flows in the opposite direction, as indicated by $\overline{\mbox{T} }$. In other word, the $\overline{\mbox{T} }$ sign means that the diagram is to the right of an (imaginary) final-state cut (that is, in the complex conjugate amplitude).

Note that each diagram corresponds to $i \, M$, such that 
\begin{align}
i \, M + (i \, M)^* = \mbox{Diagram} + \mbox{Diagram}^* . 
\end{align} 

Again, only the Im part of the amplitude contributes in \eq{B4}. \\

{\sl Flavor non-singlet case:} \\

{\bf \othree} Let us now consider the flavor non-singlet case. As follows from \eq{GdefwzNS}, the corresponding TMD is proportional to the real part of following operator expectation value \cite{Kovchegov:2016zex} (again for a quark target)
\begin{align}
\langle q | \mbox{T} \, \tr \left[ V_0 \, V_1^{pol, \, \dagger} \right] | q \rangle - \langle q | \mbox{T} \,  \tr \left[ V_1^{pol} \, V_0^\dagger \right] | q \rangle = \parbox{2.3cm}{\includegraphics[width=2.3cm]{t_quarks2}} \ - \parbox{2.3cm}{\includegraphics[width=2.3cm]{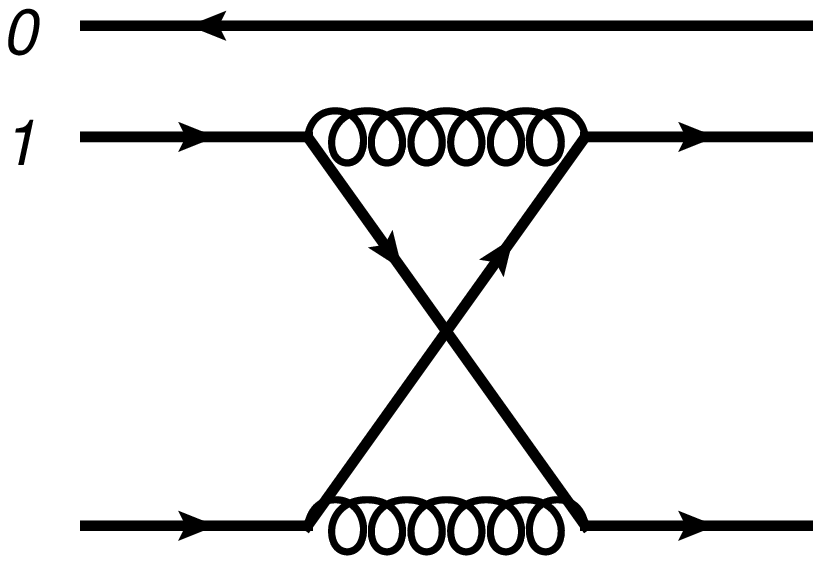}} \ .
\end{align}
Now, the second diagram contributes no imaginary part to the scattering amplitude. Its real part contribution can be related to the following diagram due to the crossing symmetry:
\begin{align}\label{B8}
 \parbox{2.3cm}{\includegraphics[width=2.3cm]{t_quarks4}} =  \ \mbox{Re} \ \parbox{2.3cm}{\includegraphics[width=2.3cm]{t_quarks2}} \ .
\end{align}
Here we assume that the corresponding amplitude in \eq{B3} is of the type $M(s, t) = f (t) \, (\ln (s/t) \pm i \, \pi)^2 + {\cal O} (1/s)$, as expected for the helicity-dependent quark exchange at the sub-eikonal level. Then the amplitude on the left of \eq{B8} is $M_{crossed} (s, t) = M (u, t)$ with $u<0$. At high energy $u \approx -s$ such that $M_{crossed} (s, t)$ = Re~$M_{crossed} (s, t)$ = Re~$M (-s, t)$ = Re~$M (s, t)$ for the leading DLA part of our ansatz for the amplitude $M(s, t)$, which is what is implied in the diagrammatic equality \eq{B8}. 

The Re sign in \eq{B8} applies to the amplitude $M$, rather than to the diagram, which is $i\, M$, such that
\begin{align}\label{B9}
 \parbox{2.3cm}{\includegraphics[width=2.3cm]{t_quarks4}} = \ \mbox{Re} \ \parbox{2.3cm}{\includegraphics[width=2.3cm]{t_quarks2}} \ = i \, \mbox{Re} M.
\end{align}
We conclude that
\begin{align}
\langle q | \mbox{T} \, \tr \left[ V_0 \, V_1^{pol, \, \dagger} \right] | q \rangle - \langle q | \mbox{T} \, \tr \left[ V_1^{pol} \, V_0^\dagger \right] | q \rangle = \parbox{2.3cm}{\includegraphics[width=2.3cm]{t_quarks2}} \ - \parbox{2.3cm}{\includegraphics[width=2.3cm]{t_quarks4}} = i \, M -  i \, \mbox{Re} M = - \mbox{Im} M. 
\end{align}
This result is in agreement with Eq.~(56) of \cite{Kovchegov:2016zex}. In addition, the expectation value of the operator from \eq{GdefwzNS} is real, thus justifying the assumption made in \cite{Kovchegov:2016zex}.

%%%%%%%%%%%%%%%%%%%%%%%%%%%%%%%%%%%%%%%%%%%%%%%%%%%%%%%%%

%||||||||||||||||||||||||||||||||||||||||||||||||||||||||||||||||||||||||||||||||||||||||||||||||||||||||||||||||||||||||
%||||||||||||||||||||||||||||||||||||||||||||||||||||||||||||||||||||||||||||||||||||||||||||||||||||||||||||||||||||||||
%||||||||||||||||||||||||||||||||||||||||||||||||||||||||||||||||||||||||||||||||||||||||||||||||||||||||||||||||||||||||

% \bibliography{references}
% \bibliographystyle{JHEP}

\providecommand{\href}[2]{#2}\begingroup\raggedright\endgroup

\end{document}